\documentclass[twocolumn]{aastex61}
\usepackage{color}
\usepackage{ulem}
\usepackage{multirow}

\shortauthors{Fukui et al.}

\begin{document}
\title{A detailed study of the interstellar protons toward the TeV $\gamma$-ray SNR RX~J0852.0$-$4622 (G266.2$-$1.2, Vela~Jr.); a third case of the $\gamma$-rays and ISM spatial correspondence}

\author{Y. Fukui}
\affiliation{Institute for Advanced Research, Nagoya University, Furo-cho, Chikusa-ku, Nagoya 464-8601, Japan; fukui@a.phys.nagoya-u.ac.jp, sano@a.phys.nagoya-u.ac.jp}
\affiliation{Department of Physics, Nagoya University, Furo-cho, Chikusa-ku, Nagoya 464-8601, Japan}

\author{H. Sano}
\affiliation{Institute for Advanced Research, Nagoya University, Furo-cho, Chikusa-ku, Nagoya 464-8601, Japan; fukui@a.phys.nagoya-u.ac.jp, sano@a.phys.nagoya-u.ac.jp}
\affiliation{Department of Physics, Nagoya University, Furo-cho, Chikusa-ku, Nagoya 464-8601, Japan}

\author{J. Sato}
\affiliation{Department of Physics, Nagoya University, Furo-cho, Chikusa-ku, Nagoya 464-8601, Japan}

\author{R. Okamoto}
\affiliation{Department of Physics, Nagoya University, Furo-cho, Chikusa-ku, Nagoya 464-8601, Japan}

\author{T. Fukuda}
\affiliation{Department of Physics, Nagoya University, Furo-cho, Chikusa-ku, Nagoya 464-8601, Japan}

\author{S. Yoshiike}
\affiliation{Department of Physics, Nagoya University, Furo-cho, Chikusa-ku, Nagoya 464-8601, Japan}

\author{K. Hayashi}
\affiliation{Department of Physics, Nagoya University, Furo-cho, Chikusa-ku, Nagoya 464-8601, Japan}

\author{K. Torii}
\affiliation{Nobeyama Radio Observatory, Minamimaki-mura, Minamisaku-gun, Nagano 384-1305, Japan}

\author{T. Hayakawa}
\affiliation{Department of Physics, Nagoya University, Furo-cho, Chikusa-ku, Nagoya 464-8601, Japan}

\author{G. Rowell}
\affiliation{School of Physical Sciences, University of Adelaide, North Terrace, Adelaide, SA 5005, Australia}

\author{M. D. Filipovi$\mathrm{\acute{c}}$}
\affiliation{Western Sydney University, Locked Bag 1797, Penrith South DC, NSW 1797, Australia}

\author{N. Maxted}
\affiliation{School of Physics, The University of New South Wales, Sydney, 2052, Australia}

\author{N. M. McClure-Griffiths}
\affiliation{Research School of Astronomy and Astrophysics, Australian National University, Canberra ACT 2611, Australia}

\author{A. Kawamura}
\affiliation{National Astronomical Observatory of Japan, Mitaka 181-8588, Japan}

\author{H. Yamamoto}
\affiliation{Department of Physics, Nagoya University, Furo-cho, Chikusa-ku, Nagoya 464-8601, Japan}

\author{T. Okuda}
\affiliation{National Astronomical Observatory of Japan, Mitaka 181-8588, Japan}

\author{N. Mizuno}
\affiliation{National Astronomical Observatory of Japan, Mitaka 181-8588, Japan}

\author{K. Tachihara}
\affiliation{Department of Physics, Nagoya University, Furo-cho, Chikusa-ku, Nagoya 464-8601, Japan}

\author{T. Onishi}
\affiliation{Department of Astrophysics, Graduate School of Science, Osaka Prefecture University, 1-1 Gakuen-cho, Naka-ku, Sakai 599-8531, Japan}

\author{A. Mizuno}
\affiliation{Institute for Space-Earth Environmental Research, Nagoya University, Chikusa-ku, Nagoya 464-8601, Japan}

\author{H. Ogawa}
\affiliation{Department of Astrophysics, Graduate School of Science, Osaka Prefecture University, 1-1 Gakuen-cho, Naka-ku, Sakai 599-8531, Japan}

\begin{abstract}
We present a new analysis of the interstellar protons toward the TeV~$\gamma$-ray SNR RX~J0852.0$-$4622 (G266.2$-$1.2, Vela~Jr.). We used the NANTEN2 $^{12}$CO($J$ = 1--0) and ATCA $\&$ Parkes H{\sc i} datasets in order to derive the molecular and atomic gas associated with the TeV~$\gamma$-ray shell of the SNR. We find that atomic gas over a velocity range from $V_\mathrm{LSR}$ = $-4$ km s$^{-1}$ to 50 km s$^{-1}$ or 60 km s$^{-1}$ is associated with the entire SNR, while molecular gas is associated with a limited portion of the SNR. The large velocity dispersion of the H{\sc i} is ascribed to the expanding motion of a few H{\sc i} shells overlapping toward the SNR but is not due to the Galactic rotation. The total masses of the associated H{\sc i} and molecular gases are estimated to be $\sim2.5 \times 10^4 $ $M_{\odot}$ and $\sim10^3$ $M_{\odot}$, respectively. A comparison with the H.E.S.S. TeV~$\gamma$-rays indicates that the interstellar protons have an average density around 100 cm$^{-3}$ and shows a good spatial correspondence with the TeV~$\gamma$-rays. The total cosmic ray proton energy is estimated to be $\sim10^{48}$ erg for the hadronic $\gamma$-ray production, which may still be an underestimate by a factor of a few due to a small filling factor of the SNR volume by the interstellar protons. This result presents a third case, after RX~J1713.7$-$3946 and HESS~J1731$-$347, of the good spatial correspondence between the TeV~$\gamma$-rays and the interstellar protons, lending further support for a hadronic component in the $\gamma$-rays from young TeV~$\gamma$-ray SNRs.
\end{abstract}


\section{Introduction} \label{sec:intro}
The origin of the cosmic rays (CRs) is one of the most fundamental issues in modern astrophysics since the discovery of the CRs by Victor Franz Hess in 1912. It is in particular important to understand the origin of the CR protons, the dominant constituent of the CRs. There have been a number of studies to address the acceleration sites of CR protons in the Galaxy \citep[e.g.,][]{1935Natur.135..617H}. It is widely accepted that CR particles are accelerated up the energies below $10^{15}$ eV via the diffusive shock acceleration (DSA), which takes place for instance between the upstream and downstream of the high velocity shock front of the SNR \citep[e.g.,][]{1978MNRAS.182..147B,1978ApJ...221L..29B}. It is crucial to verify observational signatures for the hadronic $\gamma$-rays of the CR proton origin.

The recent advent of high resolution $\gamma$-ray observations have allowed us to image some 20 SNRs, offering a new opportunity to identify the hadronic $\gamma$-rays. Among them, four SNRs, RX~J1713.7$-$3946, RX~J0852.0$-$4622, RCW~86 and HESS~J1731$-$347, show shell-type TeV $\gamma$-rays morphology as revealed by High Energy Stereoscopic System (H.E.S.S.) \citep{2004Natur.432...75A,2005A&A...437L...7A,2006A&A...449..223A,2007A&A...464..235A,2007ApJ...661..236A,2009ApJ...692.1500A,2011A&A...531A..81H,2016arXiv160104461H,2016arXiv160908671H,2016arXiv161101863H}. Potential hadronic $\gamma$-rays are produced by the proton-proton collisions followed by neutral pion decay, while any leptonic $\gamma$-rays come from CR electrons via the inverse-Compton effect and possibly Bremsstrahlung.

Previous interpretations of the $\gamma$-ray observations usually used mainly the $\gamma$-ray spectra in order to discern the above two processes, while some attempts were made to consider explicitly the target protons in the interstellar medium (ISM) \citep[e.g.,][]{2006A&A...449..223A,2007ApJ...661..236A}. A conventional assumption is that the target protons are uniform, having density of 1 cm$^{-3}$ \citep[e.g.,][]{2006A&A...449..223A}. Such a low density is preferred because the DSA works efficiently at such low density. It is however becoming recognized that the $\gamma$-ray spectrum alone may not be sufficient to settle the origin of the $\gamma$-rays because the penetration of the CR protons into the target ISM may significantly depend on the density of the ISM \citep{2007Ap&SS.309..365G}; the higher energy CR protons can penetrate deeply into the surrounding molecular cloud cores, whereas the lower energy CR protons cannot, making the $\gamma$-ray spectrum significantly harder than the fully interacting case \citep[][hereafter I12: see also references therein]{2012ApJ...744...71I}. I12 therefore suggest that the $\gamma$-ray spectra are not usable to discern the $\gamma$-ray production mechanisms, but that the spatial correspondence between the $\gamma$-rays and ISM distribution is a key element in testing for a hadronic $\gamma$-ray component. The effect of clumpy ISM was also discussed by \cite{2010ApJ...708..965Z} and \cite{2014MNRAS.445L..70G}. It is often noted that the acceleration via DSA must happen in a low-density space, which has too low a density for the hadronic origin to be effective \citep[e.g.,][]{2010ApJ...712..287E}. This is however not a difficulty if one takes into account the highly inhomogeneous ISM distribution as is commonly the case in a stellar wind evacuated cavity with a dense surrounding ISM shell like in RX~J1713.7$-$3946 \citep{2010ApJ...724...59S}.

Several of the previous studies analyzed explicitly the distribution of the candidate target ISM protons observed by CO and compared them with the $\gamma$-ray distribution in the most typical TeV $\gamma$-ray SNRs including RX~J1713.7$-$3946 and RX~J0852.0$-$4622. \cite{2006A&A...449..223A} compared the TeV $\gamma$-ray distribution with the mm-wave rotational transition of CO, the tracer of H$_2$, obtained with the NANTEN 4 m telescope. These authors found some similarity between CO and $\gamma$-rays in RX~J1713.7$-$3946, whereas part of the $\gamma$-ray shell was completely missed in CO. The authors thus did not reach a firm conclusion on the target ISM protons in the hadronic scenario. In case of RX~J0852.0$-$4622, \cite{2007ApJ...661..236A} made a similar comparison in a velocity range of 0--20 km s$^{-1}$ but found little sign of the target protons in CO. So, the ISM associated with RX~J0852.0$-$4622 remained ambiguous.

\cite{2012ApJ...746...82F} (hereafter F12) carried out a detailed analysis of the ISM protons toward the RX~J1713.7$-$3946 by employing both the molecular and atomic protons, and have shown for the first time that these ISM protons have a good spatial correspondence with the TeV $\gamma$-rays. This new study showed that the atomic protons are equally important as molecular protons as the target in the hadronic process, which was not previously taken into account. The study by F12 provides a necessary condition for the hadronic process, lending support for the hadronic scenario, although this correspondence alone does not exclude the leptonic process. F12 and I12 further considered the other relevant aspects including magneto-hydro-dynamical numerical simulations of the SN shocks and X-ray observations \citep[see also][]{2010ApJ...724...59S,2013ApJ...778...59S,2015ApJ...799..175S} and argued that the TeV $\gamma$-rays in RX~J1713.7$-$3946 is likely emitted via the hadronic process.

\begin{figure*}
\begin{center}
\includegraphics[width=\linewidth,clip]{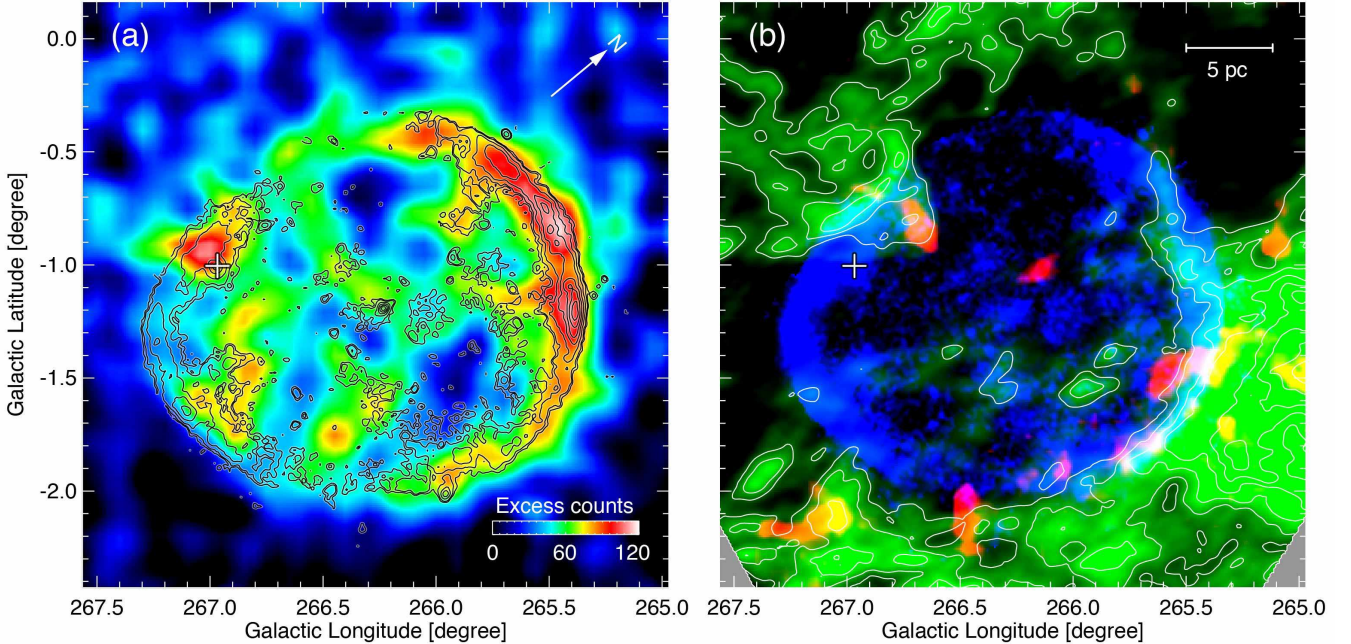}
\caption{(a) Distribution of the $\gamma$-rays and X-rays in the RX~J0852.0$-$4622 region. The TeV $\gamma$-rays in false color are obtained with H.E.S.S.~ ($E$ $>$ 250 GeV) at an angular resolution of 0.06 degrees \citep{2007ApJ...661..236A} and the black contours represent $Suzaku$ X-rays ($E$: 2.0--5.7 keV) at an angular resolution of 2 arcmin. Contour levels are 0.20, 0.25, 0.40, 0.65, 1.00, 1.45, and 2.00 $\times$ 10$^{-3}$ counts s$^{-1}$ pixel$^{-1}$. The white cross indicates the position of PSR~J0855$-$4644 \citep{2013A&A...551A...7A}. (b) Three color image of the SNR RX~J0852.0$-$4622, consisting of the NANTEN $^{12}$CO($J$ = 1--0) \citep{2001PASJ...53.1025M} in red, the ATCA $\&$ Parkes H{\sc i} in green, and the $Suzaku$ X-rays in blue. The velocity renege is from 22 km s$^{-1}$ to 33 km s$^{-1}$ for CO;  from 28 km s$^{-1}$ to 33 km s$^{-1}$ for H{\sc i}.}
\label{fig1}
\end{center}
\end{figure*}%

Another young SNR RX~J0852.0$-$4622 was discovered by \cite{1998Natur.396..141A}, and RX~J0852.0$-$4622 shows a hard X-ray spectrum toward part of the more extended Vela SNR in $ROSAT$ All-Sky Survey image. TeV $\gamma$-rays were detected and imaged toward RX~J0852.0$-$4622 by H.E.S.S.. RX~J0852.0$-$4622 shows similar properties with RX~J1713.7$-$3946; they are both young with ages of 2400--5100 yrs for RX~J0852.0$-$4622 \citep{2015ApJ...798...82A}: $\sim$1600 yr for RX~J1713.7$-$3946 \citep[e.g.,][]{2003PASJ...55L..61F,2005ApJ...631..947M} having synchrotron X-ray emission without thermal features and share shell-like X-ray/TeV $\gamma$-ray morphology. RX~J0852.0$-$4622 has an apparently large diameter of about 2 degrees. The size will allow us to test spatial correspondence between the $\gamma$-rays and the ISM at 0.12 degree angular resolution in FWHM of H.E.S.S.. We have two more shell-like TeV $\gamma$-ray SNRs RCW~86 and HESS~J1731$-$347 \citep{2009ApJ...692.1500A,2011A&A...531A..81H}. In HESS~J1731$-$347, \cite{2014ApJ...788...94F} carried out a comparative analysis of the ISM and $\gamma$-rays and have shown that the ISM protons show that their spatial distributions are similar with each other, being consistent with a dominant hadronic component of g-rays plus minor contribution of a leptonic component. In RCW~86, a similar comparative analysis is being carried out by \cite{sano2017inprep}.

\begin{figure*}
\begin{center}
\includegraphics[width=\linewidth,clip]{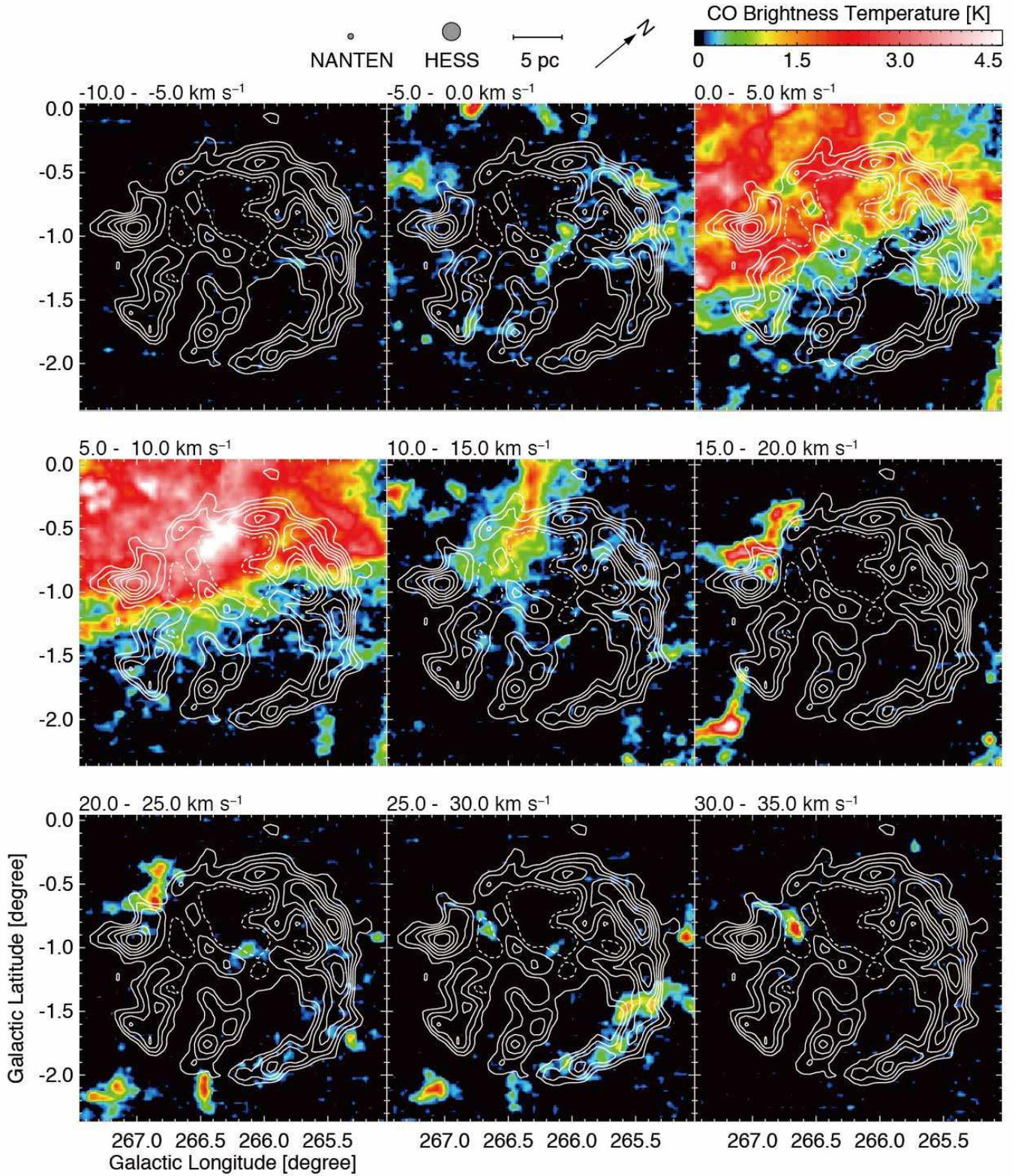}
\caption{Velocity channel distribution of $^{12}$CO($J$ = 1--0) obtained with the NANTEN telescope. Each panel shows a CO brightness temperature averaged over the velocity range from $-10$ km s$^{-1}$ to 80 km s$^{-1}$ every 10 km s$^{-1}$. Superposed are the H.E.S.S.~ $\gamma$-ray contours in white at every 10 counts from 50 counts as the lowest. The beam sizes of NANTEN and H.E.S.S.~ are shown as full width at half maximum (FWHM).}
\label{fig2}
\end{center}
\end{figure*}%

\begin{figure*}
\figurenum{2}
\begin{center}
\includegraphics[width=\linewidth,clip]{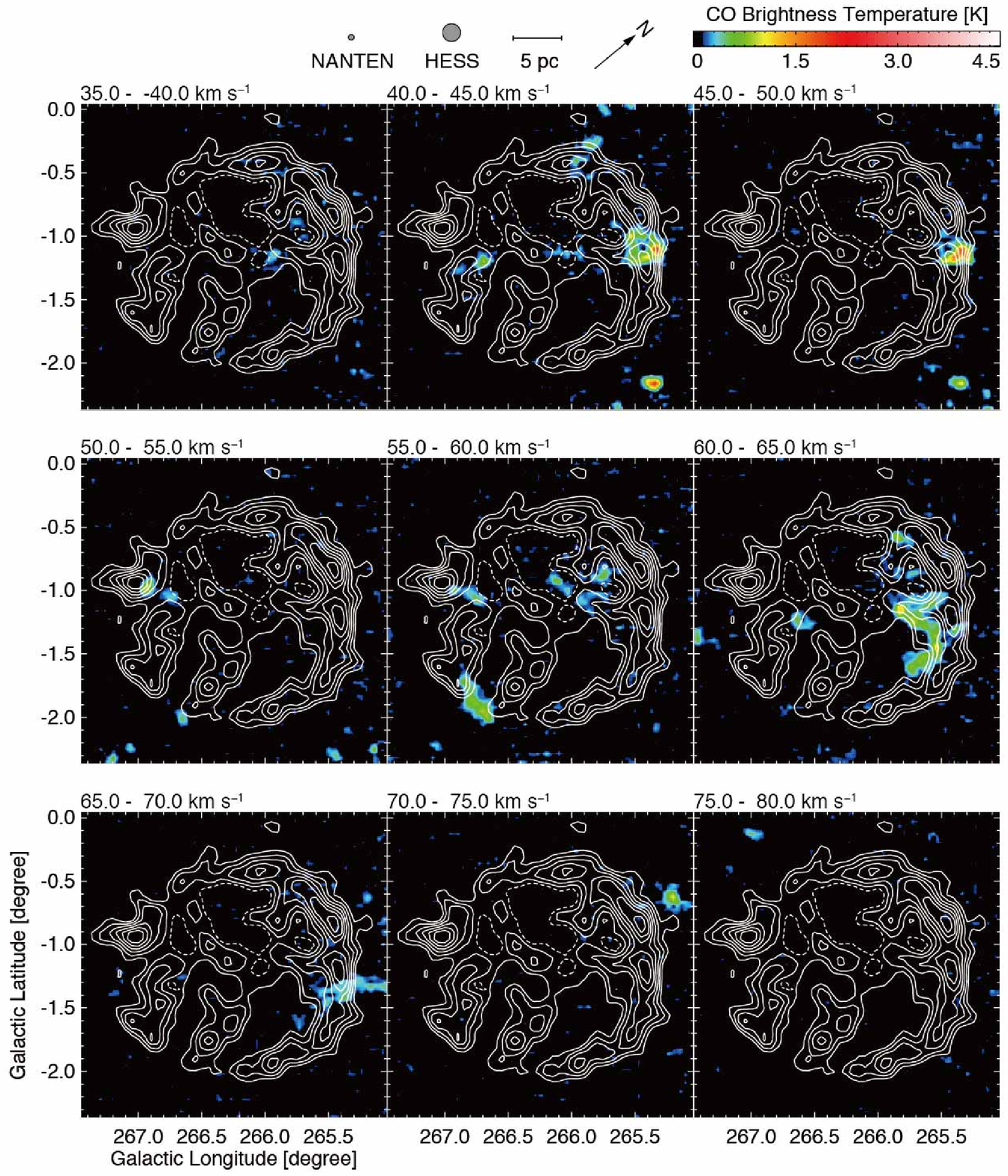}
\caption{$Continued.$}
\label{fig2}
\end{center}
\end{figure*}%

\begin{figure*}
\begin{center}
\includegraphics[width=\linewidth,clip]{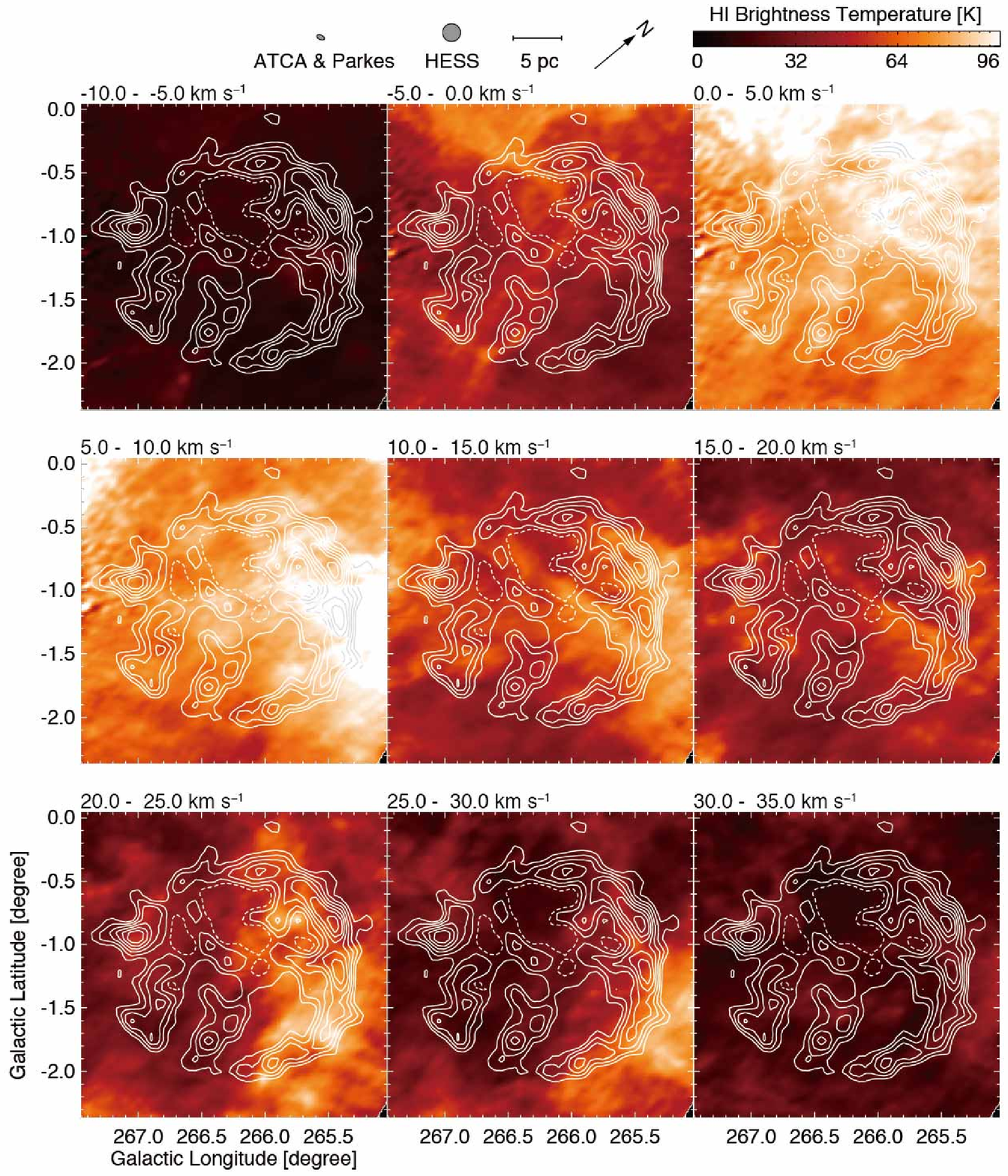}
\caption{Velocity channel distribution of H{\sc i} obtained with the ATCA $\&$ Parkes telescopes. Each panel shows the H{\sc i} brightness temperature averaged over the velocity range from $-10$ km s$^{-1}$ to 80 km s$^{-1}$ every 10 km s$^{-1}$. Superposed are the same as Figure \ref{fig2}.}
\label{fig3}
\end{center}
\end{figure*}%

\begin{figure*}
\begin{center}
\figurenum{3}
\includegraphics[width=\linewidth,clip]{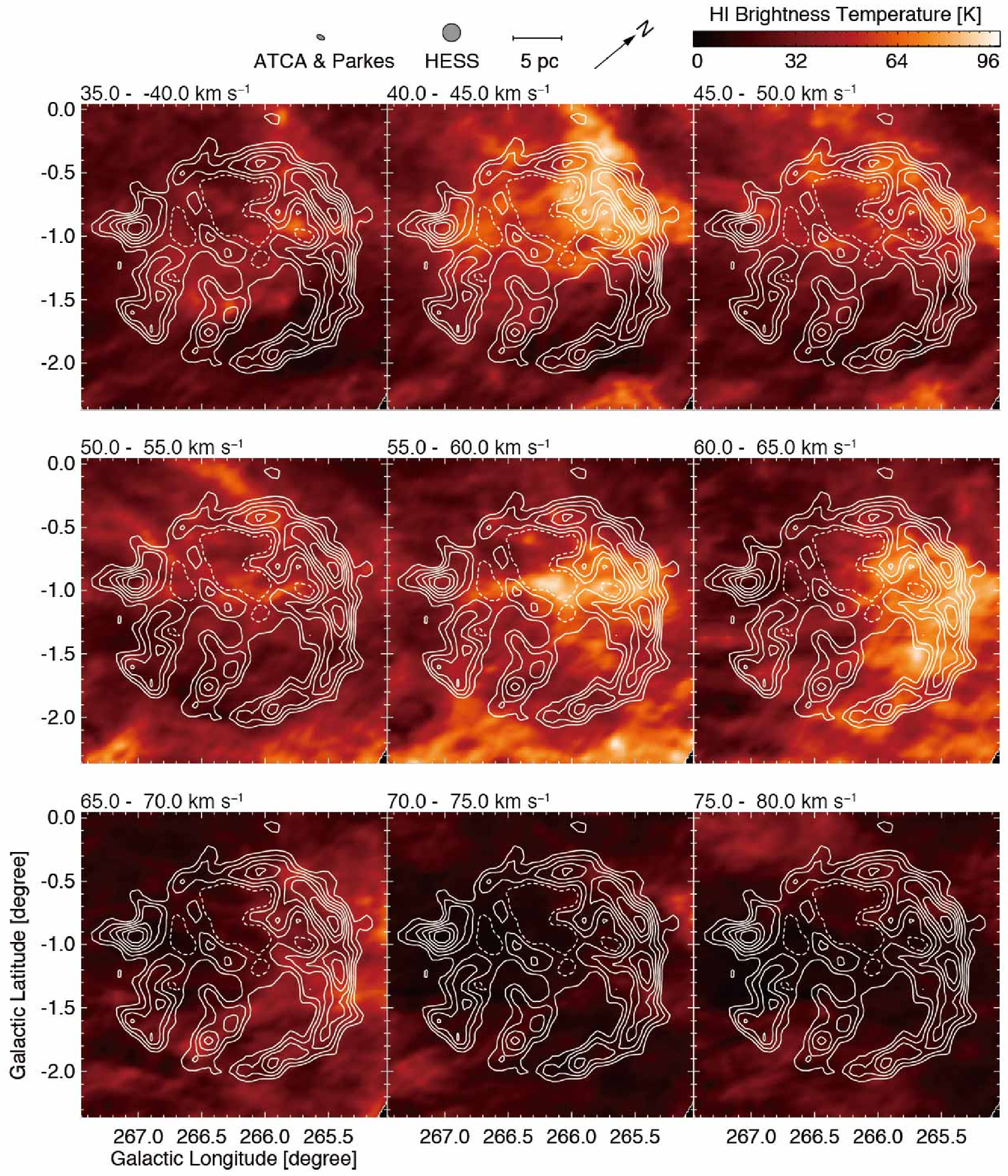}
\caption{$Continued.$}
\label{fig3}
\end{center}
\end{figure*}%

\begin{deluxetable*}{lccccccc}[t]
\tablenum{1}
\label{tab1}
\tablewidth{\linewidth}
\tablecaption{Properties of CO Clouds toward the SNR RX~J0852.0$-$4622}
\tablehead{\multicolumn{1}{c}{Name} & $l$ & $b$ & $T_{\rm R^\ast} $ & $V_{\mathrm{peak}}$ & $\Delta V$ & Size & Mass \\
& (degree) & (degree) & (K) & \scalebox{0.9}[1]{(km $\mathrm{s^{-1}}$)} & \scalebox{0.9}[1]{(km $\mathrm{s^{-1}}$)} & (pc) &  ($10^4$ $M_\sun $) \\
\multicolumn{1}{c}{(1)} & (2) & (3) & (4) & (5) & (6) & (7) & (8)}
\startdata
CO0S & 267.07 & $-1.63$ & \phantom{0}2.86 & \phantom{0}3.47 & 3.48 & 3.2 & \phantom{0}220 \\
CO20E & 266.87 & $-0.67$ & 11.46 & 19.80 & 2.06 & 6.8 & 1330 \\
CO25W & 266.47 & $-2.04$ & \phantom{0}7.48 & 24.00 & 1.72 & 6.2 & \phantom{0}650 \\
CO25C & 266.13 & $-1.00$ & \phantom{0}2.04 & 24.45 & 2.68 & 1.9 & \phantom{0}\phantom{0}40 \\
CO30E & 266.67 & $-0.87$ & \phantom{0}6.28 & 31.09 & 2.78 & 3.0 & \phantom{0}180 \\
CO45NW & 265.33 & $-1.10$ & \phantom{0}4.42 & 44.68 & 4.07 & 4.0 & \phantom{0}330 \\
CO60NW & 265.53 & $-1.47$ & \phantom{0}3.31 & 63.98 & 2.25 & 6.3 & \phantom{0}410 \\
\enddata
\tablecomments{Col. (1): Name of CO cloud. Cols. (2--7): Physica properties of the CO cloud obtained by a single Gaussian fitting. Cols. (2)--(3): Position of the peak intensity in the Galactic coordinate. Col. (4): Radiation temperature. Col. (5): Center velocity. Col. (6): Full-width half-maximum (FWHM) of line width. Col. (7): Size of CO cloud defined as $2 \times (A / \pi)^{0.5}$, where $A$ is the area of cloud surface surrounded by the CO contours in Figure \ref{fig5}. Col. (8): Mass of CO cloud is defined as $m_\mathrm{H}$ $\mu$  $\sum_{i} [D^2$ $\Omega$ $N(\mathrm{H_2})]$, where $m_\mathrm{H}$ is the mass of the atomic hydrogen, $\mu$ is the mean molecular weight, $D$ is the distance to RX~J0852.0$-$4622, $\Omega$ is the angular size in pixel, and $N$($\mathrm{H_2}$) is column density of molecular hydrogen for each pixel. We used $\mu = 2.8$ by taking into account the helium abundance of $20\%$ relative to the molecular hydrogen in mass, and $N$($\mathrm{H_2}$) = 2.0 $\times$ $10^{20}$[$W$($^{12}$CO) (K km $\mathrm{s^{-1}}$)] ($\mathrm{cm^{-2}}$) \citep{1993ApJ...416..587B}.}
\end{deluxetable*}

The distance of RX~J0852.0$-$4622 was not well determined in the previous works \cite[e.g.,][]{2001ApJ...548..814S,2001AIPC..565..403S,2007ESASP.622...91I,2010ApJ...721.1492P,2008ApJ...678L..35K,2015ApJ...798...82A,2017MNRAS...submitted}. A possible distance of RX~J0852.0$-$4622 was $250 \pm 30$ pc similar to the Vela SNR \citep{1999ApJ...515L..25C}, while another possible distance was larger than that of the Vela SNR. \cite{2001ApJ...548..814S,2001AIPC..565..403S} argued that RX~J0852.0$-$4622 is physically associated with the giant molecular cloud, the Vela Molecular Ridge \citep[VMR,][]{1988A&AS...73...51M,1999PASJ...51..775Y}, and the distance of the VMR was estimated to be 700 pc to 200 pc \citep{1992A&A...265..577L}. It is however not established if the VMR is physically associated with RX~J0852.0$-$4622 \citep[see e.g.,][]{2010ApJ...721.1492P}. Toward the northwest-rim of RX~J0852.0$-$4622, two observations were made with the $XMM$-$Newton$ and an expansion velocity was derived and an age of 1.7--4.3$\times 10^3$ years was estimated by \cite{2008ApJ...678L..35K}, assuming the free expansion phase. Recently, \cite{2015ApJ...798...82A} improved the expansion measurement by using two $Chandra$ datasets separated by 4.5 years toward the northwestern rim of RX~J0852.0$-$4622. They derived an age of 2.4--5.1$\times 10^3$ year old and the range of distance from 700 to 800 pc, which are roughly consistent with the previous study by \cite{2008ApJ...678L..35K}. We will therefore adopt the distance $\sim$750 pc to RX~J0852.0$-$4622 in the present paper. Analyses of a central X-ray source in this SNR using data from multiple observatories suggest that the progenitor of the SNR is a high-mass star which led to a core-collapse type SNe \citep{1998Natur.396..141A,2001ApJ...548..814S,2001ApJ...548L.213M,2001ApJ...559L.131P}. This conclusion is also consistent with the estimate of the current SNR shell expansion speed \citep{1999ApJ...514L.103C}.

In this paper, we present the results of an analysis of the ISM protons toward RX~J0852.0$-$4622 by using both the CO and H{\sc i} data. Section \ref{sec:obs} gives the observations of CO and H{\sc i}. Section \ref{sec:dist} describes the high-energy $\gamma$-ray and X-ray data. Section \ref{sec:ism} gives the results of the CO and H{\sc i} analyses and Section \ref{sec:discuss} discussion. We conclude the paper in Section \ref{sec:sum}.

\begin{figure*}
\begin{center}
\figurenum{4a}
\includegraphics[width=\linewidth,clip]{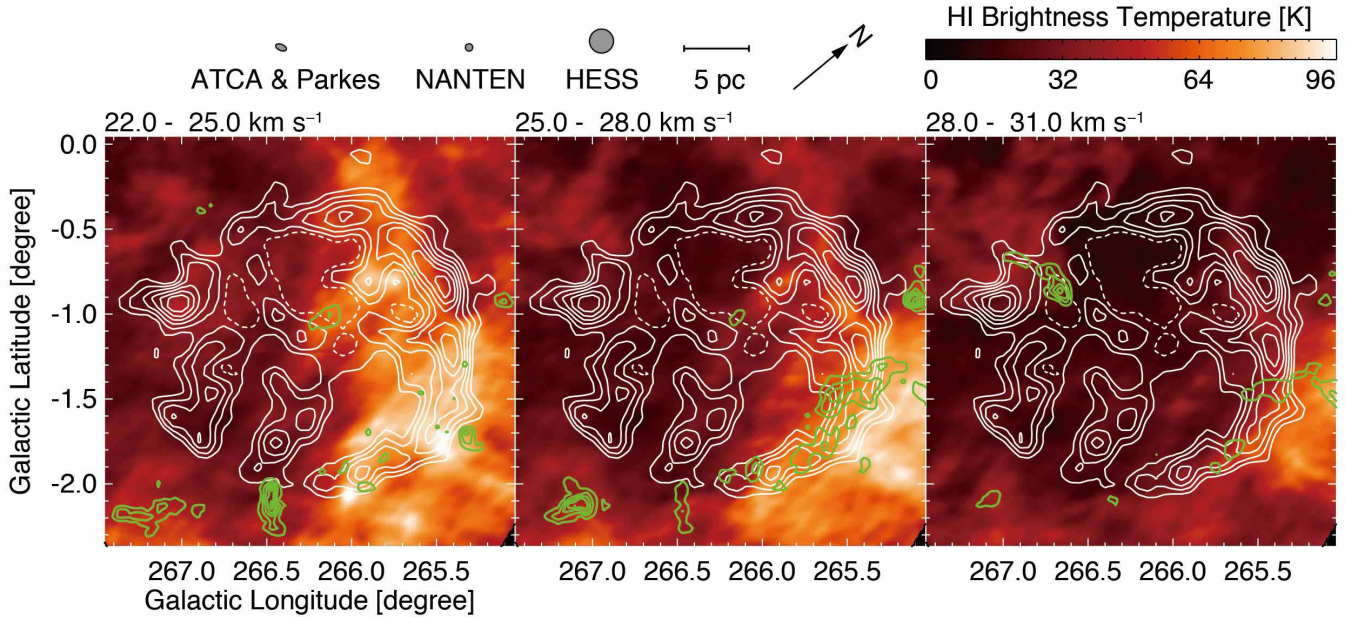}
\caption{Velocity channel distribution of the H{\sc i} images superposed with the $^{12}$CO($J$=1--0) contours toward (a) the whole region, (b) the east region, and (c) the north region. The velocity range is from 22 km s$^{-1}$ to 31 km s$^{-1}$ with 3 km s$^{-1}$ interval in (a); from 28 km s$^{-1}$ to 34 km s$^{-1}$ with 1 km s$^{-1}$ interval in (b); from 22 km s$^{-1}$ to 28 km s$^{-1}$ with 1 km s$^{-1}$ interval in (c). The CO contours are every 0.57K ($\sim6\sigma$) from 0.38 K ($\sim4\sigma$) in (a); every 0.81 K ($\sim6\sigma$) from 0.40 K ($\sim3\sigma$) in (b) and (c). White crosses and dashed circles in (b) and (c) indicate the center of the SNR ($l$, $b$) $\sim$ (266\fdg28, $-1\fdg24$) and the approximate shell boundary (radius $\sim$0.76 degrees), respectively. White contours in (a) are the same as in Figure \ref{fig2}.}
\label{fig4a}
\end{center}
\end{figure*}%

\begin{figure*}
\begin{center}
\figurenum{4b}
\includegraphics[width=\linewidth,clip]{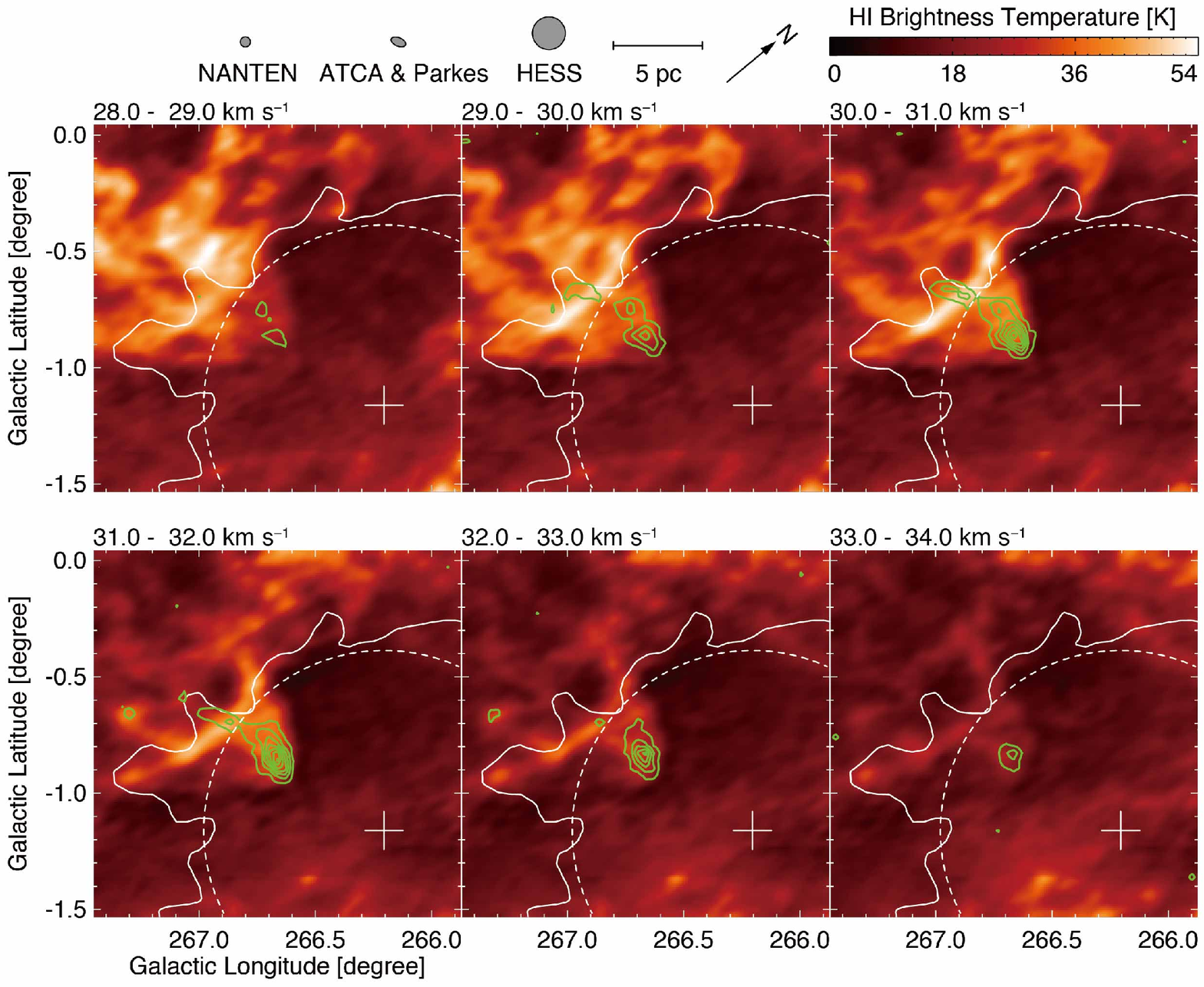}
\caption{$Continued.$}
\label{fig4b}
\end{center}
\end{figure*}%

\begin{figure*}
\begin{center}
\figurenum{4c}
\includegraphics[width=\linewidth,clip]{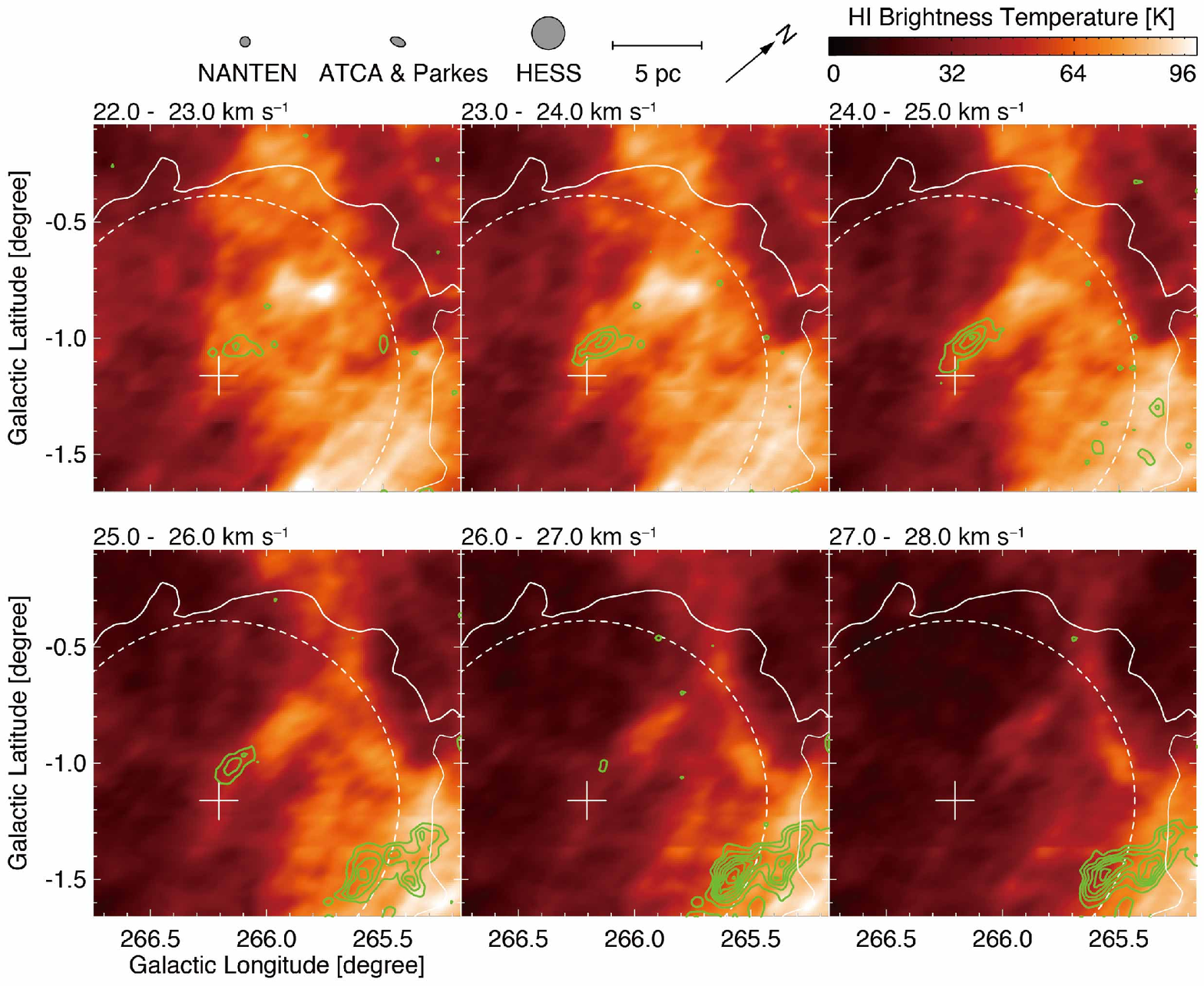}
\caption{$Continued.$}
\label{fig4c}
\end{center}
\end{figure*}%

\section{Observations} \label{sec:obs}
\subsection{CO observations}
Observations of the $^{12}$CO($J$ = 1--0) transition was carried out by the NANTEN 4 m telescope of Nagoya University at Las Campanas Observatory (2400 m above the sea level) in Chile in 1999 May--July \citep{2001PASJ...53.1025M}. The half-power beam width (HPBW) at the frequency of the $^{12}$CO($J$ = 1--0), 115.290 GHz, was $160''$. The observations were made by position switching mode and grid spacing was $120''$. SIS (superconductor-insulator-superconductor) mixer receiver provided $T_\mathrm{sys}$ of $\sim250$ K in the single side band (SSB) including the atmosphere in the direction of the zenith. The spectrometer was an AOS (acousto-optical spectrometer) with 40 MHz bandwidth and 40 kHz resolution, providing 100 km s$^{-1}$ velocity coverage and a velocity resolution of 0.1 km s$^{-1}$. The velocity always refers to that of the local standard of rest. The rms noise per channel is $\sim0.5$ K.

\subsection{H{\sc i} observations}
We used the high resolution 21 cm H{\sc i} data obtained with Australia Telescope Compact Array (ATCA) in Narrabri, New South Wales in Australia consisting of six 22 m dishes. The region of RX~J0852.0$-$4622 in $l$ = 266 degree was observed as part of the Southern Galactic Plane Survey \citep[SGPS;][]{2005ApJS..158..178M} and was combined with the H{\sc i} data taken with the 64 m Parkes telescope. The Parkes H{\sc i} data were taken for $b$ = $-10$ to $+10$ degrees, while the ATCA covers $b$ = $-1.5$ to $+1.5$ degrees. In order to supplement the southwestern half of the SNR new H{\sc i} observations were conducted during 24 hours on February 26--27 and March 29--30, 2011, with the ATCA in the EW352 and EW367 configurations (Project ID: C2449, PI: Y. Fukui). We employed the mosaicking technique, with 43 pointings arranged in a hexagonal grid at the Nyquist separation of $19'$. The absolute flux density was scaled by observing PKS B1934$-$638, which was used as the primary bandpass and amplitude calibrator. We also periodically observed PKS 0823$-$500 for phase and gain calibration. The MIRIAD software package was used for the data reduction \citep{1995ASPC...77..433S}. We combined the ATCA dataset with singledish observations taken with the Parkes 64 m telescope and the SGPS dataset. The final beam size of H{\sc i} is $245''$ $\times$ $130''$ with a position angle of 117 degrees. Typical rms noise level was 1.4 K per channel for a velocity resolution of 0.82 km s$^{-1}$.

\section{Distributions of the SNR} \label{sec:dist}
Figure \ref{fig1}a shows the TeV $\gamma$-ray and X-ray distributions of RX~J0852.0$-$4622. These high energy features are thin shell-like and are enhanced toward the northern half. 

\begin{figure}
\begin{center}
\figurenum{5}
\includegraphics[width=\linewidth,clip]{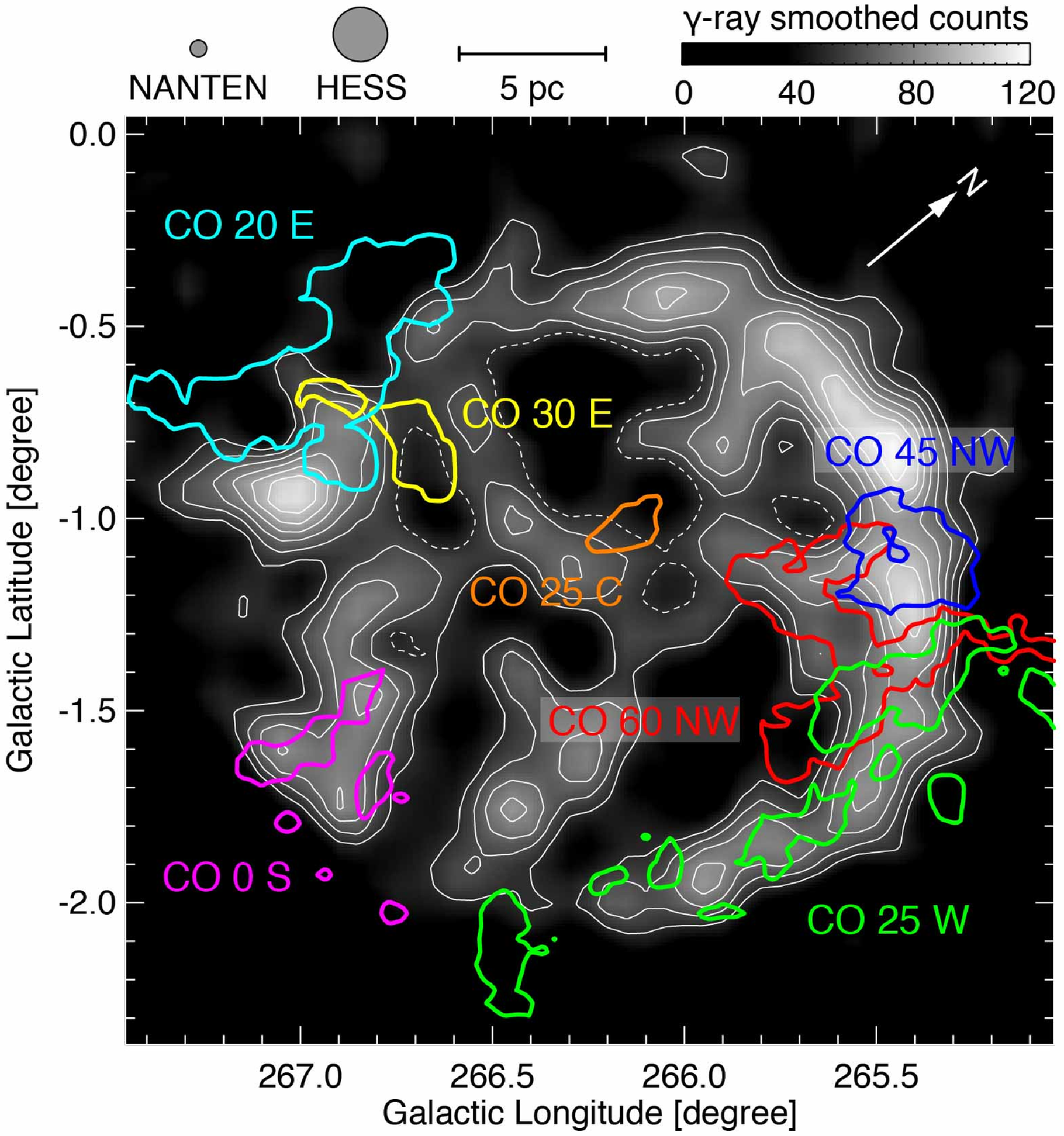}
\caption{A summary of the $^{12}$CO($J$ = 1--0) features in RX~J0852.0$-$4622. The gray image and white contours show the TeV $\gamma$-ray distribution and the color contours show the CO features. The integration velocity ranges are as follows: $-3$ km s$^{-1}$ to 7 km s$^{-1}$ (contour level: 4.4 K km s$^{-1}$, magenta) for the CO~0~S cloud; 15 km s$^{-1}$ to 24 km s$^{-1}$ (contour level: 1.8 K km s$^{-1}$, cyan) for the CO~20~E cloud; 22 km s$^{-1}$ to 30 km s$^{-1}$ (contour level: 2.3 K km s$^{-1}$, green) for the CO~25~W cloud; 22 km s$^{-1}$ to 30 km s$^{-1}$ (contour level: 1.7 K km s$^{-1}$, orange) for the CO~25~C cloud; 28 km s$^{-1}$ to 35 km s$^{-1}$ (contour level: 1.7 K km s$^{-1}$, yellow) for the CO~30~E cloud; 40 km s$^{-1}$ to 50 km s$^{-1}$ (contour level: 1.9 K km s$^{-1}$, blue) for the CO~45~NW cloud; and 61 km s$^{-1}$ to 67 km s$^{-1}$ (contour level: 1.5 K km s$^{-1}$, red) for the CO~60~NW cloud.}
\label{fig5}
\end{center}
\end{figure}%

\begin{figure*}
\begin{center}
\figurenum{6}
\includegraphics[width=\linewidth,clip]{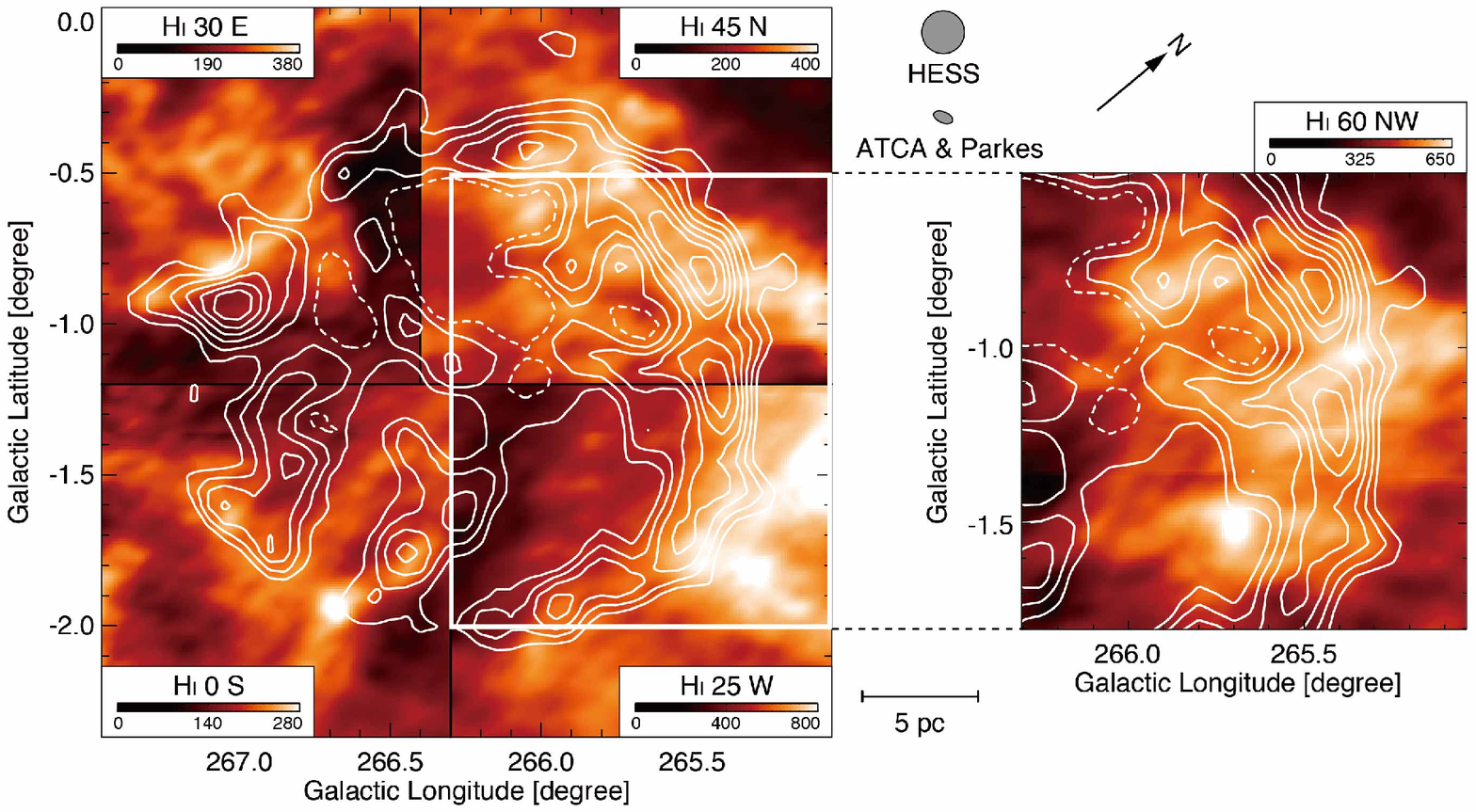}
\caption{A summary of the H{\sc i} features in RX~J0852.0$-$4622. The integration velocity ranges are as follows: $-4$ km s$^{-1}$ to 1 km s$^{-1}$ for the H{\sc i}~0~S cloud; 23 km s$^{-1}$ to 34 km s$^{-1}$ for the H{\sc i}~25~W cloud; 26 km s$^{-1}$ to 35 km s$^{-1}$ for the H{\sc i}~30~E cloud; 44 km s$^{-1}$ to 48 km s$^{-1}$ for the H{\sc i}~45~N cloud; 59 km s$^{-1}$ to 66 km s$^{-1}$ for the H{\sc i}~60~NW cloud. The white contours are the same as Figure \ref{fig2}.}
\label{fig6}
\end{center}
\end{figure*}%

The $\gamma$-ray distribution \citep[Figure \ref{fig1} of][]{2007ApJ...661..236A} is obtained in the energy above 0.3 TeV with High Energy Stereoscopic System (H.E.S.S.) installed in Namibia which utilized the first four 12 m diameter Cherenkov telescopes of the current five-telescopes H.E.S.S. array. The H.E.S.S. image has a point-spread function of 0.06 degrees at the 68 $\%$ radius (hereafter termed ``$r_{68}$'') or the half power full width of 0.144 degrees for 33 hr observations\footnote{A higher statistics $\gamma$-ray image was presented by \cite{2016arXiv161101863H} with an angular resolution of $r_{68}$ = 0.08 degree, poorer than 0.06 degree reported for the morphological analysis in \cite{2007ApJ...661..236A}. Since there is no significant difference between the two images except for the angular resolution and photon statistics, we therefore use the previous image in \cite{2007ApJ...661..236A} to improve the morphological study in this paper.}. The shell is well resolved for the 2-degree diameter, making RX~J0852.0$-$4622 suitable for testing the spatial correspondence between the ISM and the $\gamma$-rays. The $\gamma$-rays peaked toward ($l$, $b$) $\sim$ (266\fdg97, $-1\fdg00$) corresponds to a pulsar wind nebula PSR J0855$-$4644 at a distance of below 900 pc, and is not related to the RX~J0852.0$-$4622 \citep[][]{2013A&A...551A...7A}. The total $\gamma$-ray flux in the energy range from 0.3--30 TeV is estimated to be (84.1 $\pm$ $4.3_\mathrm{stat}$ $\pm$ $21.7_\mathrm{syst}$) $\times$ $10^{-12}$ erg cm$^{-2}$ s$^{-1}$ \citep{2016arXiv161101863H}.

$ROSAT$ observations \citep{1998Natur.396..141A,1999A&A...350..997A}, $ASCA$ observations \citep{2000PASJ...52..887T,2001ApJ...548..814S,2001AIPC..565..403S}, and $Suzaku$ observations \citep{2016PASJ...68S..10T} show that the object is shell-like, with luminous northern and southern rims and less luminous northeastern and southwestern rims. The X-rays are superposed on the $\gamma$-ray image in Figure \ref{fig1}a. The most prominent X-ray peak is seen toward the northwestern shell, ($l$ ,$b$) = (265\fdg4, $-1\fdg2$) \citep{2000PASJ...52..887T,2001ApJ...548..814S,2001AIPC..565..403S}. The X-rays are non-thermal with a photon index around 2.7 and the absorbing column density of 2--4$\times 10^{21}$ cm$^{-2}$ \citep[e.g.,][]{2000PASJ...52..887T,2001ApJ...548..814S,2001AIPC..565..403S,2005ApJ...632..294B,2005A&A...429..225I}. $ASCA$ and $Chandra$ observations of the SNR show that the non-thermal X-rays are dominant with an absorption column density 1--4$\times10^{21}$ cm$^{-2}$ toward the X-ray peak. A spatially-resolved spectroscopic X-ray study revealed details of the fine structure in the luminous northwestern rim complex of G~266.2$-$1.2 by observations made with $Chandra$ \citep{2005ApJ...632..294B,2005A&A...429..225I,2010ApJ...721.1492P,2015ApJ...798...82A}.

\begin{figure*}
\begin{center}
\figurenum{7}
\includegraphics[width=\linewidth,clip]{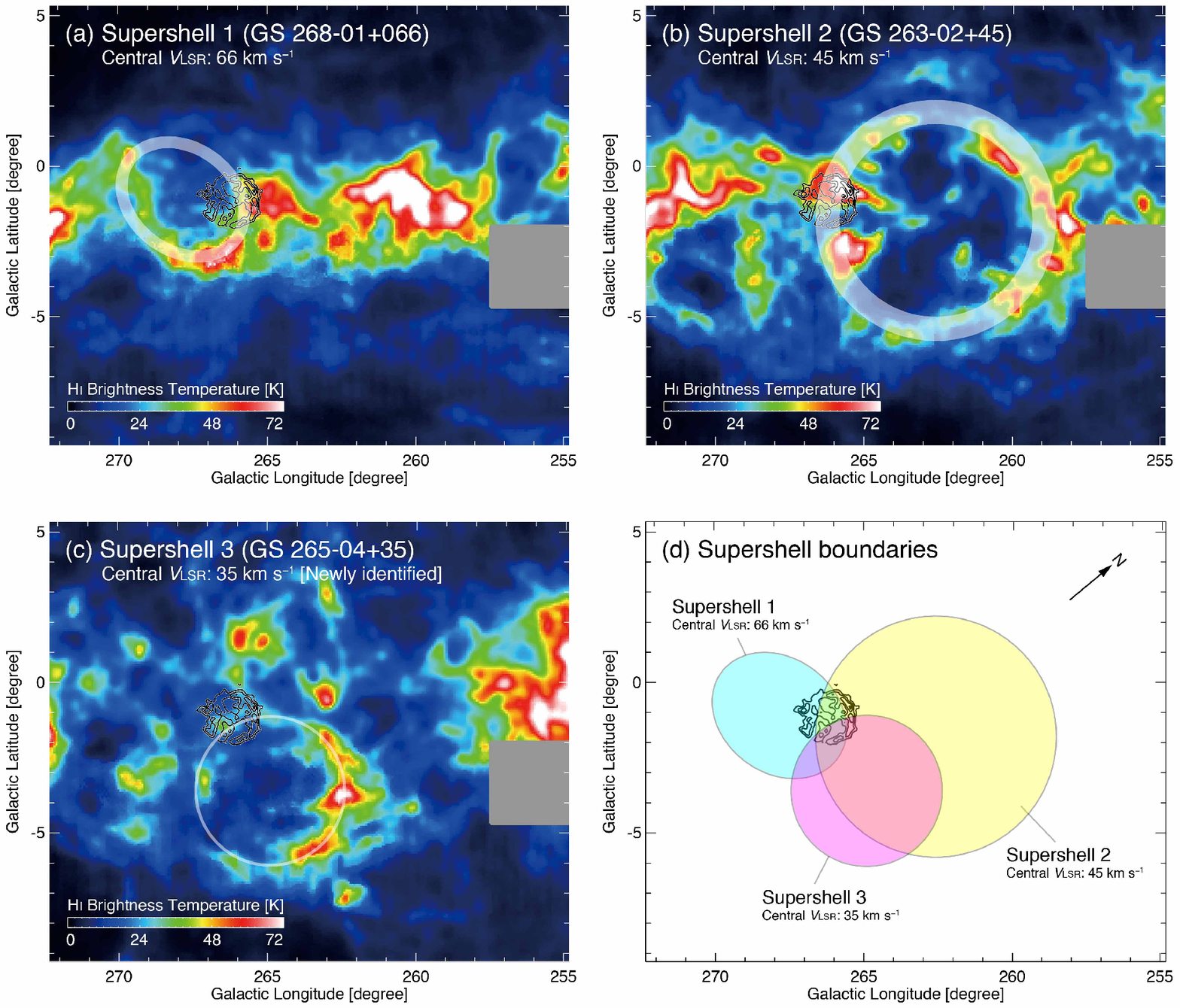}
\caption{H{\sc i} supershells toward the region of RX~J0852.0$-$4622. H{\sc i} intensity maps of supershell 1 \citep[a: GS~268$-$01$+$066;][]{2014AA...564A.116S}, supershell 2 \citep[b: GS~263$-$02$+$45;][]{2007AA...476..255A}, and supershell 3 (c: GS~265$-$04$+$35; This work), respectively. The central velocities are shown in each panel. The white circles indicate the best-fit radius having thickness of an error. Superposed are the H.E.S.S.~ $\gamma$-ray contours in white at every 20 counts from 50 counts as the lowest. (d) Schematic view of the supershell boundaries toward the SNR RX~J0852.0$-$4622. Each circle shows the outer boundaries of the H{\sc i} supershells. The detailed physical parameters are shown in Table \ref{tab2} (see also Appendix).}
\label{fig7}
\end{center}
\end{figure*}%

\section{ISM distributions} \label{sec:ism}
\subsection{CO and H{\sc i} distributions}
CO is a tracer of molecular gas with density of a few times 100 cm$^{-3}$ or higher and the H{\sc i} traces the atomic gas with lower density less than several 100 cm$^{-3}$. The density range between them may be probed by the cool H{\sc i} gas as seen in self-absorption if the background H{\sc i} is bright enough. Such H{\sc i} self-absorption is in fact observed in RX~J1713.7$-$3946 and is identified as part of the target ISM protons in the hadronic $\gamma$-ray production (F12).

Figure \ref{fig1}b shows a three-color image of RX~J0852.0$-$4622, consisting of $^{12}$CO($J$ = 1--0) in red, H{\sc i} in green, and X-ray (2--5.7 keV) in blue. H{\sc i} emission delineates the outer boundary of the X-ray shell, except for the southeast and northeast. In the western region, the X-ray shell is well correlated with the CO clumps.

Figures \ref{fig2} and \ref{fig3} show the velocity channel distributions of $^{12}$CO($J$ = 1--0) and H{\sc i} every 5 km s$^{-1}$ in the range from $-10$ km s$^{-1}$ to 80 km s$^{-1}$, where the H.E.S.S. TeV $\gamma$-ray contours are superposed. The giant molecular cloud at 0--15 km s$^{-1}$ is the VMR \citep{1991A&A...247..202M,1999PASJ...51..765Y,1999PASJ...51..775Y,2001PASJ...53.1025M}. The other CO features are all small and clumpy. We list the observational parameters of the small CO clouds in Table \ref{tab1}; the position of peak intensity within the cloud, velocity, line intensity, size, and mass.

\begin{deluxetable*}{lcccccccccc}
\tablenum{2}
\label{tab2}
\tablecaption{Observed Properties of H{\sc i} Supershells toward RX J0852.0$-$4622}
\tablehead{
\multicolumn{1}{c}{ID} & Name & $l$ & $b$ & $v_\mathrm{c}$ & $r$ & $a$ & $b$ & $\phi$ & $v_\mathrm{exp}$ & Reference \\
& & (deg) & (deg) & (km $\mathrm{s^{-1}}$) & (deg) & (deg) & (deg) & (deg) &  (km $\mathrm{s^{-1}}$) &\\
\multicolumn{1}{c}{(1)} & (2) & (3) & (4) & (5) & (6) & (7) & (8) & (9) & (10) & (11)}
\startdata
1 & GS 268$-$01$+$066 & 267.8 & $-1.1$ & 66 & ----- & 2.2$\pm$0.2 & 1.7$\pm$0.2  & $-36.9$ & 10.3 & [1]\\
2 & GS 263$-$02$+$45\phantom{0} & 262.6 & $-1.8$ & 45 & 3.6\phantom{0}$\pm$0.4\phantom{0} & ----- & ----- & ----- & 14\phantom{0} & [2]\\
3 & GS 265$-$04$+$35\phantom{0} & 264.9 & $-3.6$ & 35 & 2.47$\pm$0.05 &  ----- & ----- & ----- & 9 & This work\\
\enddata
\tablecomments{Col. (1): Supershell ID. Col. (2): Supershell name. Cols. (3--4): Central position of the the supershell. Col. (5): Central radial velocity of the supershell. Col. (6): Radius of the fitted circle. Cols. (7--8): Major and minor semi-axis of the fitted ellipse. Col. (9): Major axis inclination $\phi$ relative to the Galactic longitude, measured counterclockwise form the Galactic plane. Col. (10) Velocity extension of the supershell. Col. (11): [1] \cite{2014AA...564A.116S}, [2] \cite{2007AA...476..255A}.}
\end{deluxetable*}

\subsection{Associated clouds}
We shall first identify seven candidates CO clouds so named CO~0~S, CO~20~E, CO~25~W, CO~25~C, CO~30~E, CO~45~NW, and CO~60~NW, where their typical velocity is indicated by the Figure \ref{fig5}. In addition we show five H{\sc i} features, H{\sc i}~0~S, H{\sc i}~25~W, H{\sc i}~30~E, H{\sc i}~45~N, and H{\sc i}~60~NW in Figure \ref{fig6}, which are candidates for the H{\sc i} counterparts of the CO. By combining these CO with the H{\sc i} and the shell distribution, we select the plausible candidates for the clouds interacting with the SNR.

Figure \ref{fig4a} shows an overlay between CO and H{\sc i}. The CO corresponds well the south-western rim of the shell. This shows a clear association of CO cloud CO~25~W with the SNR. H{\sc i}~25~W is also associated with this CO showing a shift in position from the center to northwest in Figure \ref{fig4a}. The shift is consistent with part of an expanding shell on the near side at an expanding velocity around 5--10 km s$^{-1}$. We also note that H{\sc i}~25~W is likely located toward the tangential edge of the shell, suggesting that the central velocity of the ISM associated with the SNR is around 25 km s$^{-1}$.

Figures \ref{fig4b} and \ref{fig4c} show two additional cases of the association. Figure \ref{fig4b} shows CO~30~E with H{\sc i} tail extending outward from the centre, H{\sc i}~30~E. The H{\sc i} is V-shaped pointing toward the centre of the SNR. CO~30~E is located at the tip of the H{\sc i} and is elongated in radial direction. We suggest that the distribution presents a blown-off cloud overtaken by the stellar wind by the SN progenitor, where the dense head of CO survived against the stellar wind with a H{\sc i} tail. A similar morphology is seen toward CO~25~C in Figure \ref{fig4c}. The CO is also elongated toward the centre and the H{\sc i}, part of H{\sc i}~45~N, shows elongation in the same direction with the CO at the inner H{\sc i} tip (Figure \ref{fig4c}). We discuss more quantitative details of the interaction in the next section. The three CO features, CO~25~W and CO~25~C, in a velocity range from 20 km s$^{-1}$ to 30 km s$^{-1}$ suggest the physical association of the CO and H{\sc i} with RX~J0852.0$-$4622. We also note that the small double feature CO~0~S is seen toward the southeast of the $\gamma$-ray feature. These features and the H{\sc i}~0~S (Figure \ref{fig6}) show correspondence with the $\gamma$-rays. To summarize, the CO in a velocity range from $-3$ km s$^{-1}$ to 30 km s$^{-1}$  with a central velocity 25 km s$^{-1}$ shows signs of the association with the SNR.

\begin{figure}
\begin{center}
\figurenum{8}
\includegraphics[width=\linewidth,clip]{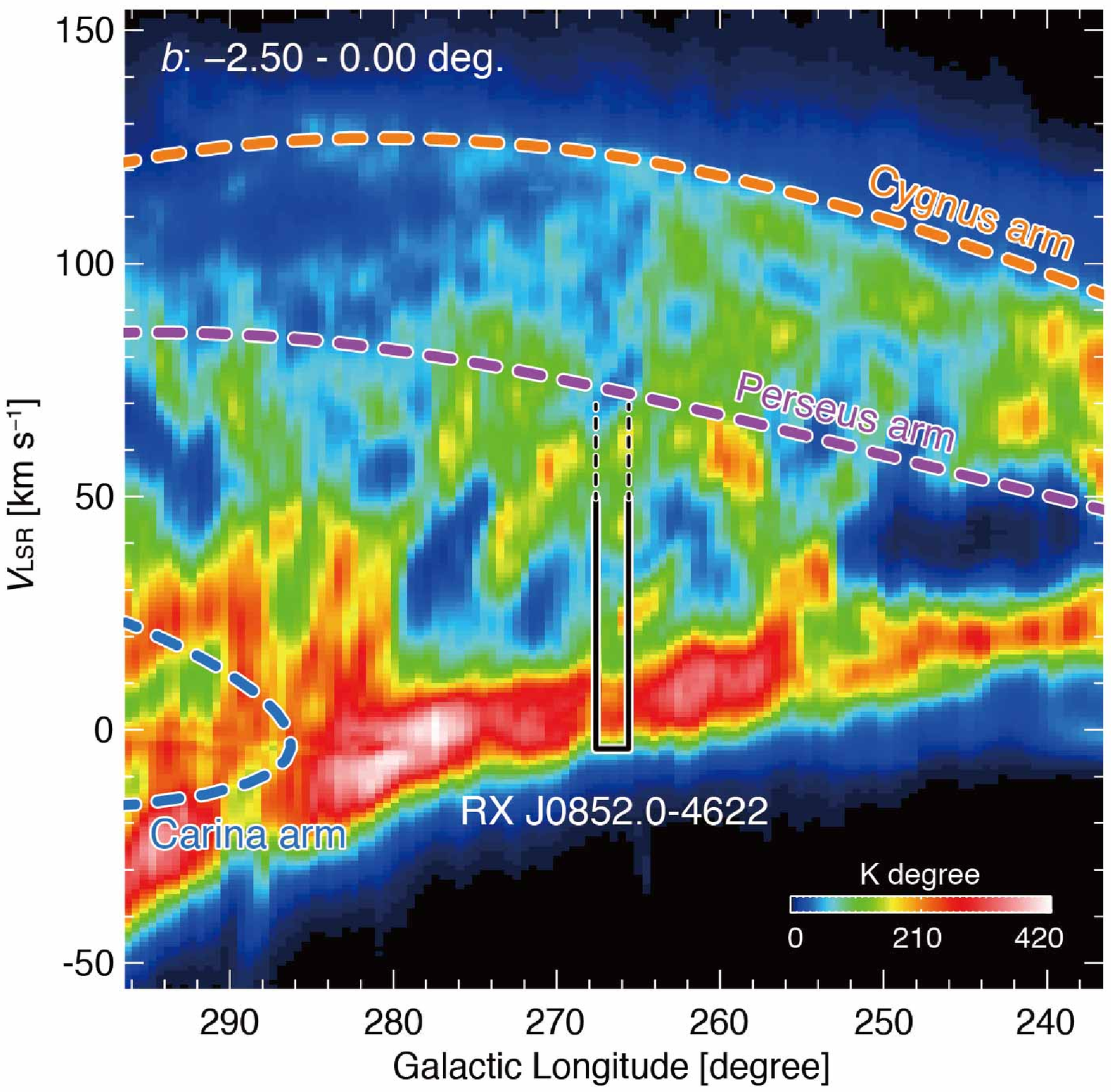}
\clearpage
\caption{Longitude-velocity diagram of the H{\sc i} emission obtained with the Leiden/Argentine/Bonn (LAB) survey \citep{2005A&A...440..775K}. The integration range is from $-2.5$ degree to 0 degree in the Galactic latitude. Black solid and bashed lines indicate the region of RX~J0852.0$-$4622. Orange, purple, and cyan dashed lines show the Cygnus arm, Perseus arm, and Carina arm, respectively \citep{2008AJ....135.1301V}.}
\label{fig8}
\end{center}
\end{figure}%

\begin{figure}
\begin{center}
\figurenum{9}
\includegraphics[width=\linewidth,clip]{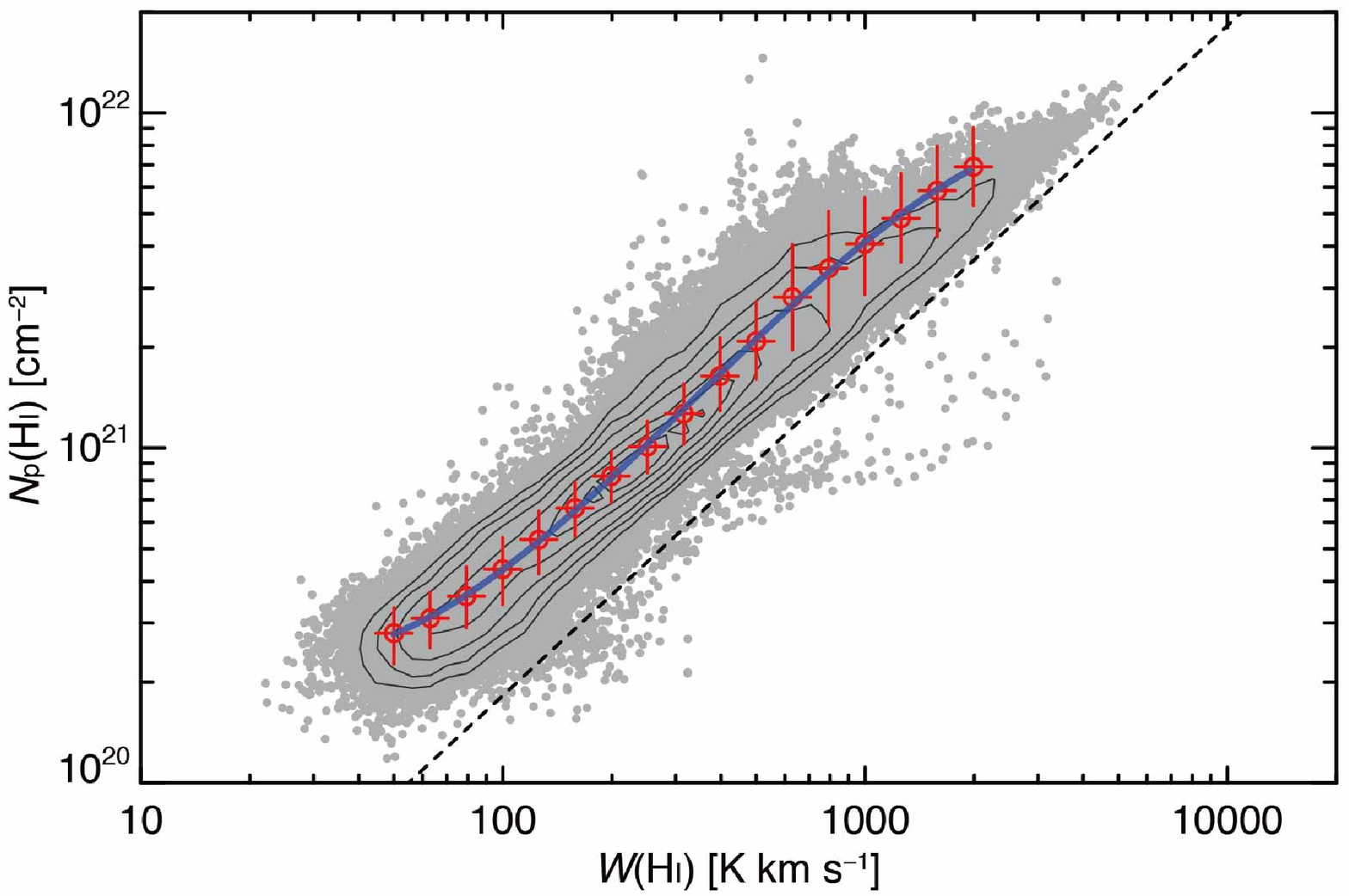}
\clearpage
\caption{Correlation plot between the $W_\mathrm{p}$(H{\sc i}) and $N_\mathrm{p}$(H{\sc i}) in consideration of the optically thick H{\sc i} (see the text for details). Red crosses and blue solid line represent the averaged values of $N_\mathrm{p}$(H{\sc i}) and their best-fit curve. The dashed line indicates the $N_\mathrm{p}$(H{\sc i}) with optically thin case.}
\label{fig9}
\end{center}
\end{figure}%

Based on these candidates for the association, we extended a search for CO and H{\sc i} in a velocity from $-5$ km s$^{-1}$ to 65 km s$^{-1}$ in Figures \ref{fig2} and \ref{fig3}, and have summarized the possible candidate features in Figures \ref{fig5} and \ref{fig6}. These candidate CO and H{\sc i} features are CO~20~E, CO~45~NW and CO~60~NW, and H{\sc i}~45~N, and H{\sc i}~60~NW. It is notable that CO~25~W is found toward the $\gamma$-ray shell (panel 25--30 km s$^{-1}$).

\subsection{Distance of the ISM: H{\sc i} supershells}
The central velocity of the ISM associated with RX~J0852.0$-$4622 is around 25 km s$^{-1}$ with a velocity span of $\sim50$ km s$^{-1}$. 25 km s$^{-1}$ corresponds to a kinematic distance of 4.3 kpc if the galactic rotation model is adopted \citep{1993A&A...275...67B}, significantly different from the adopted distance value of 750 pc \citep{2008ApJ...678L..35K,2015ApJ...798...82A}. We shall look at the H{\sc i} distribution around the SNR in order to clarify this discrepancy.

\begin{figure*}
\begin{center}
\figurenum{10}
\includegraphics[width=\linewidth,clip]{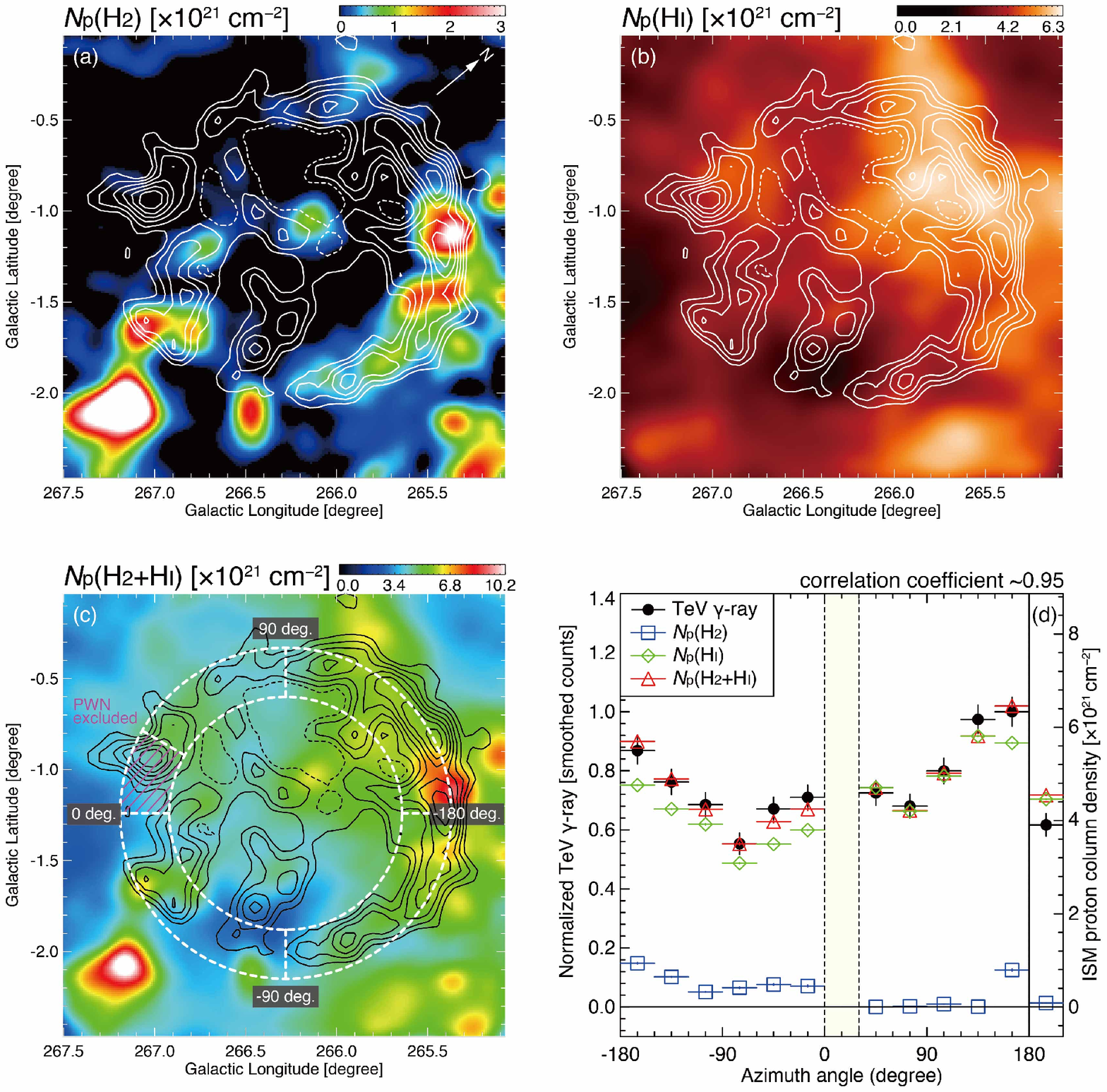}
\caption{Distributions of column density of the ISM protons (a) $N_\mathrm{p}$(H$_2$), (b) $N_\mathrm{p}$(H{\sc i}), and (c) $N_\mathrm{p}$(H$_2$ + H{\sc i}). Superposed contours are the same as in Figure \ref{fig2}. (d) Azimuthal distributions of $N_\mathrm{p}$(H$_2$), $N_\mathrm{p}$(H{\sc i}), $N_\mathrm{p}$(H$_2$ + H{\sc i}), and normalized TeV $\gamma$-ray between the two elliptical rings shown in (c). The proton column densities are averaged between the rings in units of cm$^{-2}$ (see the text). The same plots inside the inner ring are shown on the right side of (c). The shaded area was excluded due to the contamination of the PWN (see the text).}
\label{fig10}
\end{center}
\end{figure*}%

Figures \ref{fig7}a, \ref{fig7}b, and \ref{fig7}c show the H{\sc i} distributions in the velocities of 66 km s$^{-1}$, 45 km s$^{-1}$, and 35 km s$^{-1}$, respectively. By a conventional Galactic rotation model, these velocities correspond to distances of 8.7 kpc, 6.4 kpc, and 5.4 kpc \citep{1993A&A...275...67B}. We argue here that the Galactic rotation is not the dominant cause of the H{\sc i} velocity but that the expanding motion of several H{\sc i} supershells is mainly responsible for the line of sight velocity. In fact, \cite{2014AA...564A.116S} and \cite{2007AA...476..255A} identified the H{\sc i} supershells, GS~268$-$01$+$066 and GS~263$-$02$+$45 (see Figures \ref{fig7}a and \ref{fig7}b) toward the SNR. In addition, we newly identified the H{\sc i} supershell ``GS~265$-$04$+$35" as shown in Figure \ref{fig7}c. The supershell has also an expanding motion and shows the front and rear wall in the H{\sc i} spectrum (see more details in Appendix). Figure \ref{fig7}d shows a schematic view of the supershell boundaries toward the SNR RX~J0852.0$-$4622. The TeV $\gamma$-ray contours are overlapped three supershells, indicating that the expanding motion of the supershells dominates the H{\sc i} velocity field around the SNR. Their detailed physical parameters are shown in Table \ref{tab2}.

The H{\sc i} position-velocity diagram on a large scale is shown in Figure \ref{fig8}. The H{\sc i} gas toward the SNR is not ordered in the spiral arm at 60 km s$^{-1}$ to 80 km s$^{-1}$ and the H{\sc i} has several velocity features at 0 km s$^{-1}$, 20--30 km s$^{-1}$ and 40--50 km s$^{-1}$ toward RX~J0852.0$-$4622 in $l$ = 263--268 degrees, all of which show kinematic properties consistent with part of an expanding shell (Figure \ref{fig7}). We therefore suggest that the apparent velocity shifts of the H{\sc i} in the region are due to the expansion of supershells which are driven by several stellar clusters located in the centre of each shell. The known molecular supershells, including the Carina flare supershell, show an expanding velocity of 10 km s$^{-1}$ with a $\sim10$ Myrs age \citep{1999PASJ...51..751F,2001PASJ...53.1003M,2008MNRAS.387...31D,2008PASJ...60.1297D,2011ApJ...728..127D,2011ApJ...741...85D,2015ApJ...799...64D}. We generally find no direct observational hints for any stellar clusters although it is not unusual that we see no clear signs of clusters, which should become faint, in 10 Myrs due to the evolution of high mass stars in the cluster.

\subsection{Total ISM protons}
We here derive the total ISM proton density. The molecular column density is calculated by canonical conversion factors and the error corresponds to 3 sigma noise fluctuations. The total intensity of $^{12}$CO($J$ = 1--0) can be converted into the molecular column density $N$(H$_2$) (cm$^{-2}$) by the following relation;

\begin{eqnarray}
N(\mathrm{H}_2) = X\mathrm{CO} \times W\mathrm{(^{12}CO)}
\label{eq1}
\end{eqnarray}

where the $X\mathrm{CO}$ factor $N$(H$_2$) (cm$^{-2}$)/$W$($^{12}$CO) (K km s$^{-1}$) is adopted as $X\mathrm{CO} = 2.0 \times 10^{20}$ (cm$^{-2}$ / K km s$^{-1}$) \citep{1993ApJ...416..587B}. Then, the total proton density is given as $N_\mathrm{p}$(H$_2$) = $2N$(H$_2$).

Usually, the atomic proton column density is estimated by assuming that the 21 cm H{\sc i} line is optically thin. If this approximation is valid, the H{\sc i} column density $N_\mathrm{p}$(H{\sc i}) (cm$^{-2}$) is estimated as follows \citep{1990ARA&A..28..215D};

\begin{eqnarray}
N_\mathrm{p}(\mathrm{H}{\textsc{i}}) = 1.823 \times 10^{18}  \int T_\mathrm{b} dV \phantom{0}(\mathrm{cm}^{-2}),  
\label{eq2}
\end{eqnarray}

where $T_\mathrm{b}$ (K) and $V$ (km s$^{-1}$) are the 21 cm intensity and the peak velocity. \cite{2014ApJ...796...59F,2015ApJ...798....6F} made a new analysis by using the sub-mm dust optical depth derived by the $Planck$ satellite \citep{2014A&A...571A..11P} and concluded that the H{\sc i} emission is generally optically thick with optical depth of around 1 in the local interstellar medium within 200 pc of the sun at the Galactic latitude higher than 15 deg. This optical depth correction increases the H{\sc i} density by a factor of $\sim2$ as compared with the optically thin case. We adopt this correction by using the relationship between the H{\sc i} integrated intensity and the 353 GHz dust optical depth $\tau_{353}$ given in Figure \ref{fig9} \citep[see also][]{2014A&A...571A..11P}. Since RX~J0852.0$-$4622 is close to the Galactic plane, the 353 GHz dust optical depth is not available for a single velocity component as in the local space. Instead, we adopt the empirical relationship between $W(\mathrm{H}{\textsc{i}})$ and $\tau_{353}$ in Figure \ref{fig9} in order to estimate the $N_\mathrm{p}(\mathrm{H}{\textsc{i}})$ from $W(\mathrm{H}{\textsc{i}})$. The ratio of the $N_\mathrm{p}(\mathrm{H}{\textsc{i}})$ and  $\tau_{353}$ is determined to as follows in the optically thin limit \citep[][equation (3)]{2015ApJ...798....6F},

\begin{eqnarray}
\frac{N_\mathrm{p}(\mathrm{H}{\textsc{i}}) }{1 \times 10^{21}  \phantom{0}(\mathrm{cm}^{-2})} = \Biggl(\frac{\tau_{353}}{4.77 \times 10^{-6}}\Biggr)^{1/1.3}
\label{eq3}
\end{eqnarray}

hence,

\begin{eqnarray}
N_\mathrm{p}(\mathrm{H}{\textsc{i}}) = (1.2 \times 10^{25}) \times (\tau_{353})^{1/1.3} \phantom{0}(\mathrm{cm}^{-2})
\label{eq3-2}
\end{eqnarray}

where the non-linear dust property, $N_\mathrm{p}(\mathrm{H}{\textsc{i}})$ $\propto$ $(\tau_{353})^{1/1.3}$ derived by \cite{2013ApJ...763...55R,2017ApJ...838..132O} is assumed, and this non linearity does not alter $N_\mathrm{p}(\mathrm{H}{\textsc{i}})$ significantly in the present $W(\mathrm{H}{\textsc{i}})$ range.

The H{\sc i} velocity range is taken to be 20 km s$^{-1}$ to 50 km s$^{-1}$ and $-4$ km s$^{-1}$ to 1 km s$^{-1}$, where VMR is dominant in the velocity range from $-5$ km s$^{-1}$ to 15 km s$^{-1}$. The CO~0~S, CO~25~W, and CO~25~W locations are also taken into account as the molecular components. The total ISM proton column density is given by the sum as follows;

\begin{eqnarray}
N_\mathrm{p}(\mathrm{H}_2 + \mathrm{H}{\textsc{i}}) = N_\mathrm{p}(\mathrm{H}_2) + N_\mathrm{p}(\mathrm{H}{\textsc{i}}),  
\label{eq4}
\end{eqnarray}

Figures \ref{fig10}a, \ref{fig10}b, and \ref{fig10}c present the total molecular protons, atomic protons and the sum of the molecular and atomic protons, respectively. The resolution is adjusted to the major axis of the beam size ($245''$).

Our assumed distance around 1 kpc for most of the associated ISM features is independently confirmed by the visual extinction $A_\mathrm{V}$ seen in the southern half of the SNR where the foreground contamination by the VMR is not significant. Figure \ref{fig12} shows the Av distribution from the Digitized Sky Survey I \citep[DSS I;][]{2005PASJ...57S...1D} where $A_\mathrm{V}$ is estimated by star counting of the 2MASS data. The direction ($l$, $b$) = (266$^{\circ}$, $-$1$^{\circ}$) is where the number of stars is small as compared with the Galactic centre and $A_\mathrm{V}$ is not very accurately estimated in general. Nevertheless we see a clear indication of enhanced $A_\mathrm{V}$ features in Figure \ref{fig12}, where $A_\mathrm{V}$ is typically 0.7 mag to 1.7 mag as is clearly seen toward the southern part of the $\gamma$-ray shell, and the CO~0~S, H{\sc i}~0~S, and CO~25~W clouds. The peaks have total proton column density of 5--10$ \times 10^{21}$ cm$^{-2}$ in Figure \ref{fig10}c, corresponding to $A_\mathrm{V}$ = 3--5 mag ($A_\mathrm{V}$ = $N_\mathrm{p}(\mathrm{H}) / 1.7 \times 10^{21}$ cm$^{-2}$ mag). The fact that the extinction is visible toward the SNR indicates that the cloud is relatively close to the sun, and the image in Figure \ref{fig12} is consistent with a distance around 1 kpc.

\subsection{TeV $\gamma$-rays}
\begin{figure}
\begin{center}
\figurenum{11}
\includegraphics[width=\linewidth,clip]{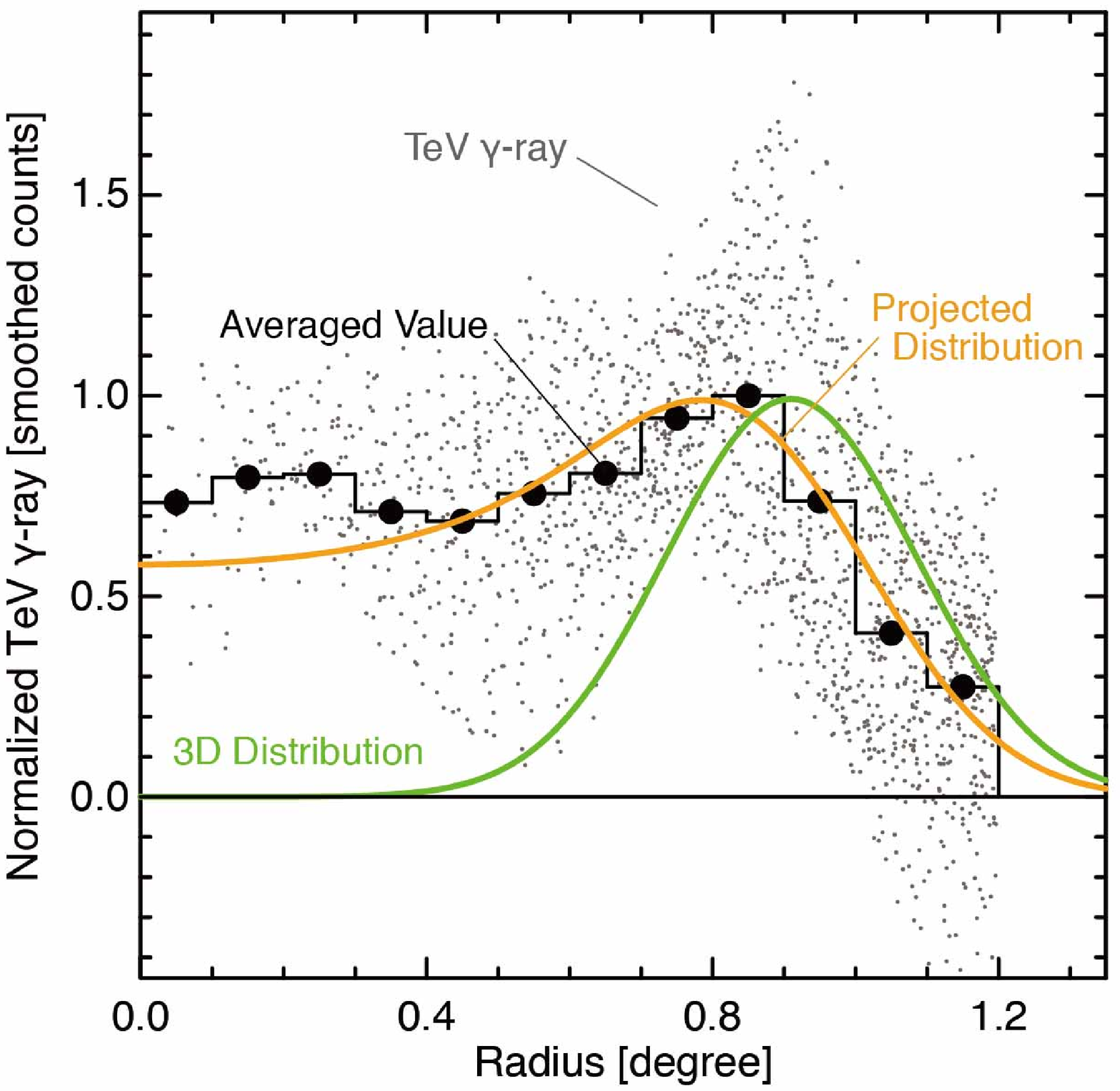}
\caption{The radial distribution of the TeV $\gamma$-rays. The normalized $\gamma$-ray smoothed counts are plotted vs. radius measured from ($l$, $b$)=(266\fdg28, $-1\fdg24$). The small dots indicate the data points and the large dots averaged values in every 0.1 degrees in radius. The three dimensional TeV $\gamma$-ray emission is fitted by a spherical shell with Gaussian intensity distribution. Fitting is made within 1.2 degrees in radius. The green line indicates the three dimensional distribution and the orange the projection on the sky. The Gaussian is peaked at 0.91 degrees ($\sim11.8$ pc) with a FWHM of 0.42 degrees ($\sim5.6$ pc). We excluded the data at an azimuthal angle from 0 degree to 30 degree due to the contamination of the PWN.}
\label{fig11}
\end{center}
\end{figure}

\begin{figure*}
\begin{center}
\figurenum{12}
\includegraphics[width=\linewidth,clip]{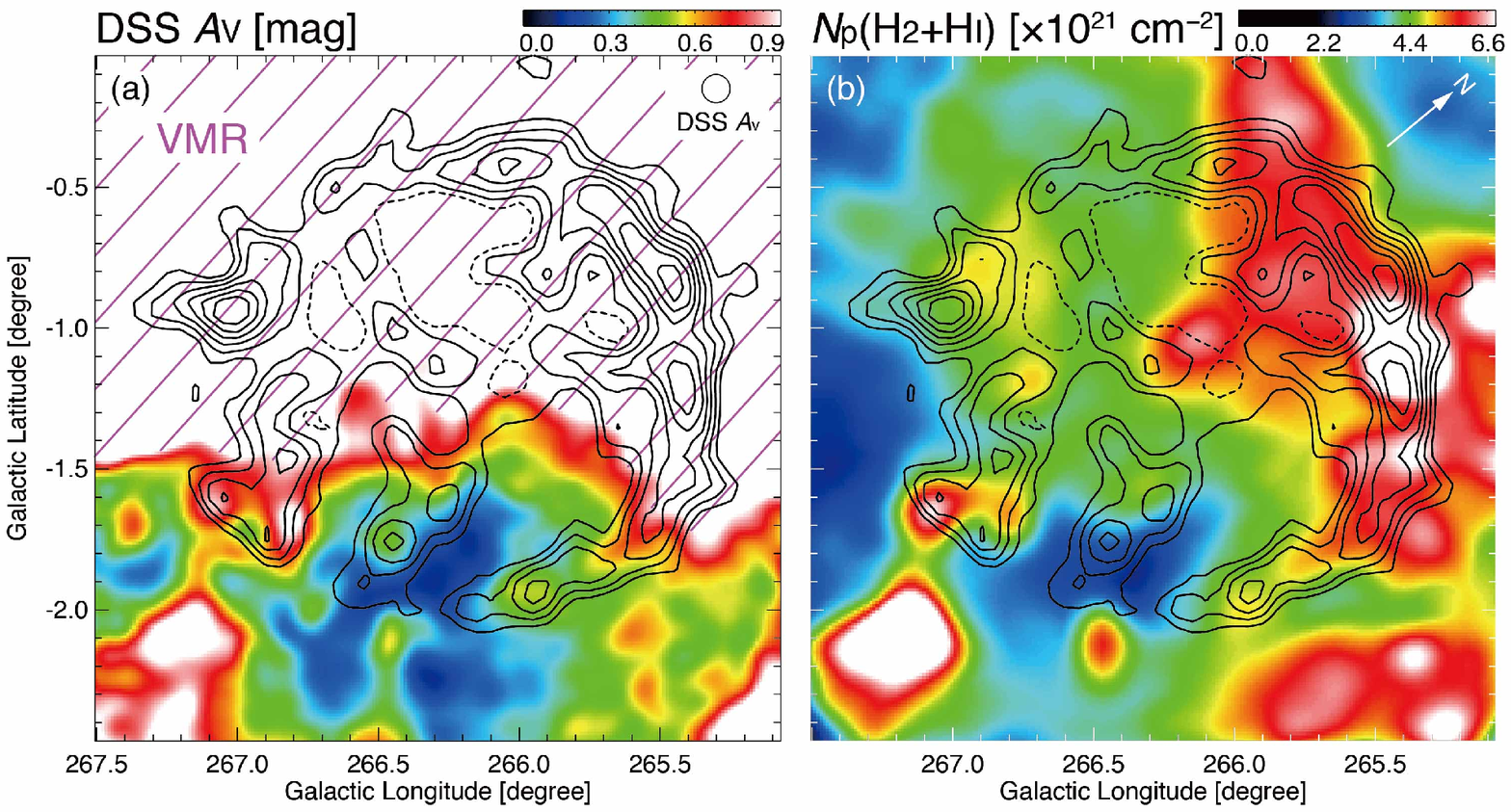}
\caption{(a) Distribution of visual extinction estimated by the Digitized Sky Survey I \citep{2005PASJ...57S...1D}. False color is the extinction and contours the TeV $\gamma$-rays. The northern half (magenta shaded) is the region of Vela Molecular Ridge. (b) Distributions of column density of the total ISM protons $N_\mathrm{p}$(H$_2$ + H{\sc i}) shown in Figure 9c. The contours are the same as in (a)}
\label{fig12}
\end{center}
\end{figure*}

The histogram in Figure \ref{fig11} indicates the average TeV $\gamma$-ray count taken every 0.1 degrees as a function of radius from the shell centre ($l$, $b$) = (266\fdg28, $-1\fdg24$) obtained by H.E.S.S. \citep{2007ApJ...661..236A}. The error is conservatively estimated as the (oversampling-corrected total smoothed count)$^{0.5}$. We shall here analyze the $\gamma$-ray distribution by using a spherical shell model. The distribution is fit by a spherically symmetric shell, following the method of F12 and a Gaussian intensity distribution in radius of the $\gamma$-ray counts as follows;

\begin{eqnarray}
F(r) =  A \times  e^{-\bigl(r - r_0\bigr)^2 / 2\sigma ^2}
\label{eq5}
\end{eqnarray}
where $A$ is a factor for normalization, the radius $r_0$ at the peak and the width of the shell $\sigma$. The green line in Figure \ref{fig11} shows the results. $r_0 = 0.91$ degrees, $\sigma = 0.18$ degrees. They correspond to $\sim12$ pc and $\sim2.4$ pc, at 750 pc.

The azimuthal $\gamma$-ray distributions are obtained by taking the two circles centered at ($l$, $b$) = (266\fdg28, $-1\fdg24$), where the radii of these circles are determined to be 0.64 degree and 0.91 degree, respectively, at the $1/3$ count level of the peak of the Gaussian shell shown in Figure \ref{fig11}. The projected distribution is shown by the orange color. The range of the radius is the same as adopted by \cite{2007ApJ...661..236A}. The fitting seems adequate in a range from 0.0 to 1.2 degrees.

\section{Discussion} \label{sec:discuss}
Spatial correspondence between $\gamma$-rays and ISM protons is a key to discerning the $\gamma$-ray production mechanism (I12, F12). Figures \ref{fig10} shows a comparison between $\gamma$-rays and ISM in the azimuthal distribution (the angle is measured clockwise with 0 degrees in the southeast). The TeV $\gamma$-ray counts were normalized.

The correspondence between the two is remarkable good, while some small deviations between them are seen at angles of $-60$ to 0 degrees and 120 to 150 degrees, and at the inside of the shell. This offers a third case after RX~J1713.7$-$3946 (F12), which supports a hadronic origin of the $\gamma$-rays via spatial correspondence. The total H{\sc i} and H$_2$ masses involved in the shell are estimated to be $\sim 2.5 \times 10^4$ $M_\odot$ and $\sim 10^3$ $M_\odot$ within the radius of $\sim15$ pc, respectively.

In Figure \ref{fig10}a there is a point which shows significant deviation at an angle of 0 to 30 degrees and this corresponds to the position of the PWN unrelated to the SNR shell, because the pulsar characteristic age is 140 kyr and the large velocity $\sim 3000$ km s$^{-1}$ needed if the pulsar is a remnant of the SNR \citep[e.g.,][]{2013A&A...551A...7A,2016arXiv161101863H}. The positions between 30 degrees and 150 degrees show a trend that the ISM proton column density under-estimates the $\gamma$-rays by less than 10 \%. These positions show no CO emission and the H{\sc i} only is responsible for the ISM protons.

In Figure \ref{fig10}c, the shell seems more intense in the $\gamma$-rays and may suggest that CR is enhanced as suggested by the significantly enhanced X-rays. We estimate the total energy of CR protons above 1 GeV by using the equation below \citep{2016arXiv161101863H};

\begin{eqnarray}
W_\mathrm{p,tot} \sim (7.1 \pm 0.3_\mathrm{stat} \pm 1.9_\mathrm{syst}) \times 10^{49} (d / 750 \;\mathrm{pc})^2 \nonumber \\
(n/1 \;\mathrm{cm}^{-3})^{-1}  \;\mathrm{(erg)} \;\;
\label{eq6}
\end{eqnarray}
where $d$ is the distance, 750 pc, and $n$ the proton density, $\sim100$ cm$^{-3}$, by adopting the shell radius $\sim$15 pc and the thickness $\sim$9 pc. $W_\mathrm{p,tot}$ estimated to be $\sim 10^{48}$ erg, corresponding to 0.1 \% of the total kinetic energy of a SNe $\sim10^{51}$ erg. The H{\sc i} distribution seems to be fairly uneven in space and velocity as suggested by the non-uniform distribution in Figure \ref{fig6}. The coupling between CR protons and the target protons may not be complete. According to \cite{2012ApJ...744...71I}, the effective mean target density for CR protons $n_\mathrm{tg}$ can be written as $n_\mathrm{tg}$ $\simeq$ $n$$f$, where $f$ is the volume filling factor of the interstellar protons. The value $W_\mathrm{p,tot}$ is therefore to be regarded as a lower limit.

In RX~J1713.7$-$3946 the total molecular mass and atomic mass are $\sim10^4$ $M_\odot$ for each and the total energy of the CR protons of 10$^{48}$ erg (F12). In RX~J0852.0$-$4622, the total atomic mass is $\sim2.5 \times 10^4$ $M_\odot$, while the total molecular mass is $\sim 10^3$ $M_\odot$. Given the radii of the two SNRs, the average density over the whole volume of RX~J1713.7$-$3946 is similar to that in RX~J0852.0$-$4622. We find that these two young TeV $\gamma$-ray SNR have total CR proton energy at an order of $10^{48}$ erg. The energy is significantly smaller than those discussed before \citep[e.g.,][]{2006A&A...449..223A} and suggests that the fraction of the explosion energy converted into CRs appears fairly low 0.1 \% in such a young stage coupled with effect of the volume filling factor. The efficiency may possibly grow in time and this can be tested by exploring middle-aged SNRs with an age of $10^4$ yrs. In fact for the middle-aged SNRs W44, W28, and IC443, the total CR proton energy is $\sim10^{49}$ erg \citep{2010A&A...516L..11G,2011ApJ...742L..30G,2013ApJ...768..179Y,2017ApJ...submitted}.


\begin{figure}
\begin{center}
\figurenum{A1}
\includegraphics[width=\linewidth,clip]{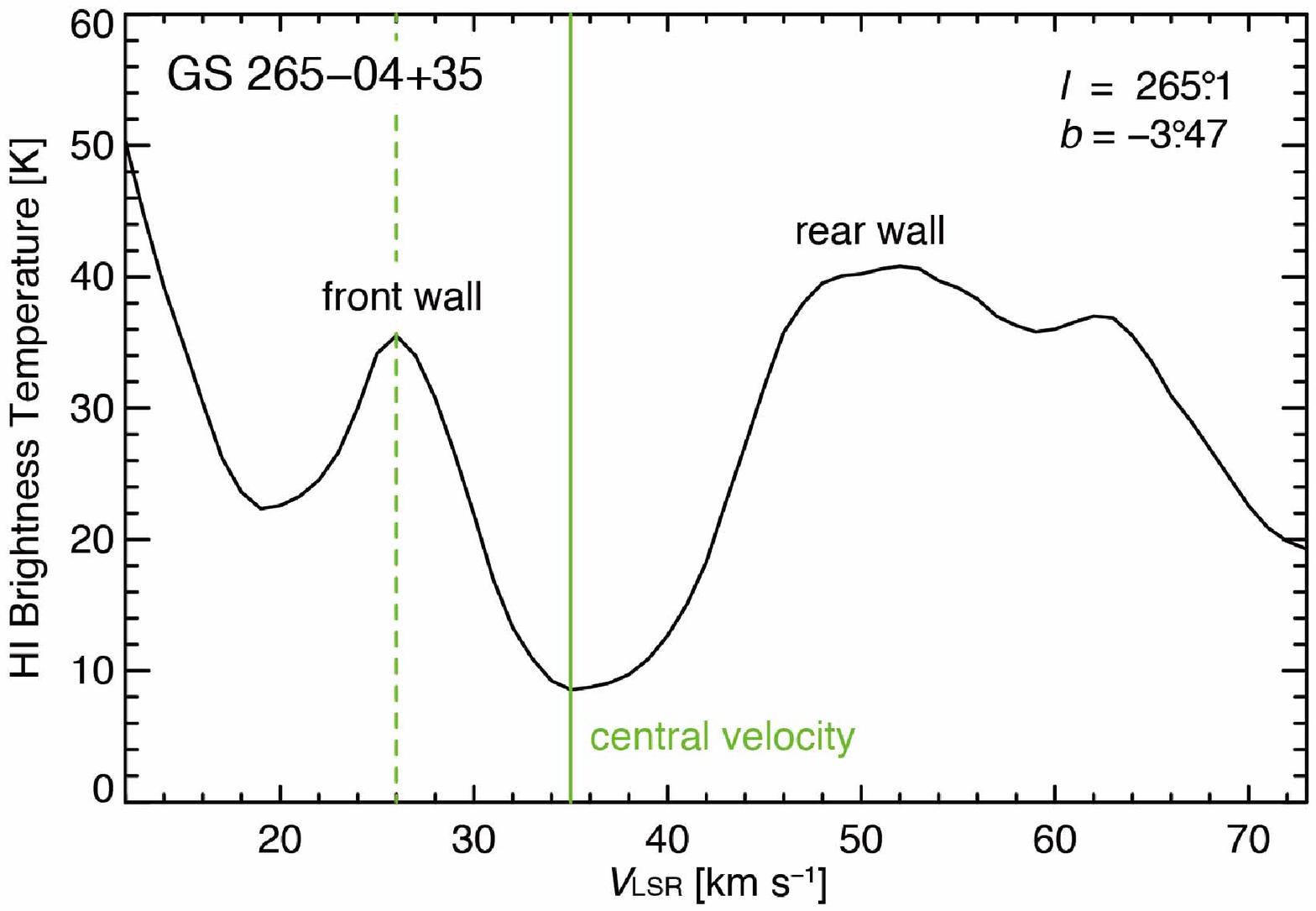}
\clearpage
\caption{Velocity profile through the supershell 3 (GS~265$-$04$+$35). The solid line corresponds to the central velocity and dashed line indicates the velocity of front wall.}
\label{figa1}
\end{center}
\end{figure}%

\begin{figure}
\begin{center}
\figurenum{A2}
\includegraphics[width=\linewidth,clip]{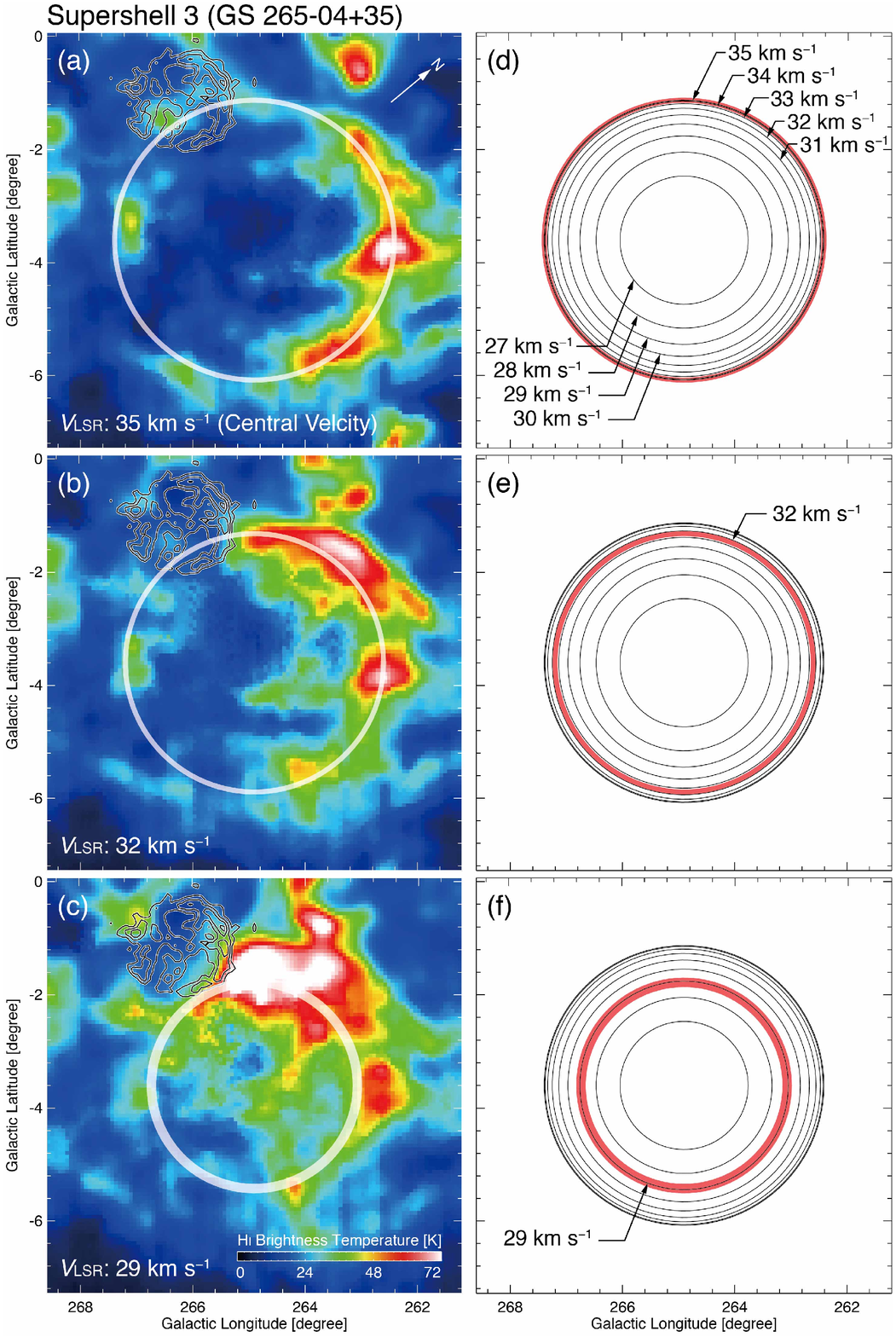}
\clearpage
\caption{(a--c) Velocity channel distributions of the H{\sc i} supershell 3 (GS~265$-$04$+$35). The boundary of fitted shells are shown by white circles for each radial velocity. (d--e) Model velocity distributions of the expanding shell of the supershell 3. Iso-velocity lines are shown here, and red areas show the corresponding velocity range shown in Figures A2(a--c).}
\label{figa2}
\end{center}
\end{figure}%

Non-thermal X-rays indicate particle acceleration in the SN blast waves. \cite{2001ApJ...548..814S,2001AIPC..565..403S} analyzed the X-rays with $ASCA$ observations. They showed that the non-thermal X-rays are dominant similarly to RX~J1713.7$-$3946. The absorption column density in front of the X-rays is $1.4 \times 10^{21}$ cm$^{-2}$ toward the western rim is consistent with the present ISM proton distribution in Figure \ref{fig10}c. ISM inhomogeneity is suggested by the CO clumps in Figure \ref{fig5} (CO~30~E and CO~25~C) overtaken by the blast waves. The situation is similar to the inside of RX~J 1713.7$-$3946. DSA works for CR acceleration in the low-density cavity and the CR protons can collide with the ISM protons in the SNR. The diffusion length is estimated to be 1 pc for the magnetic field of 10 $\mu$G and CR proton energy of 100 TeV, large enough for the CR protons to interact with the CO and H{\sc i}.

We note that there is a correlation between the $\gamma$-rays, X-rays and the ISM. All these components are enhanced on the western rim of the SNR but the eastern rim is quite weak. This distribution is interpreted in terms of the enhanced turbulence in the western rim such as suggested by magneto-hydro-dynamical numerical simulations by I12. These authors showed that the shock interacting with dense clumps create turbulence and that the turbulence amplifies the magnetic field up to 1\,mG. Thus enhanced magnetic field can intensify the synchrotron X-rays and efficiently accelerate CRs, leading to the increase of the $\gamma$- and X-rays.

RX~J0852.0$-$4622 is suggested to be a core collapse SNR \citep{1998Natur.396..141A}. The parent cloud of the SN progenitor is likely the one at 25 km s$^{-1}$ elongated along the plane by 100 pc to the west. It is the most massive H{\sc i} complex in the region. The total H{\sc i} mass of the cloud is roughly estimated to be $10^5$ $M_\odot$ and the H$_2$ mass $10^4$ $M_\odot$. The H{\sc i} cloud is found to be located in the intersection among at least three H{\sc i} supershells which are expanding at 10--15 km s$^{-1}$. Collisions among the shells may be responsible for formation of the parent cloud in this relatively diffuse environment outside the solar circle \citep[e.g.,][]{2011ApJ...728..127D,2011ApJ...741...85D}.

\section{Summary} \label{sec:sum}
We have carried out a combined analysis of CO and H{\sc i} toward the youngest TeV $\gamma$-ray SNR RX~J0852.0$-$4622 following the first good spatial correlation found in RX~J1713.7$-$3946. The main conclusions of the present study are summarized as follows;

\begin{enumerate}
\item The ISM in a velocity range from $-4$ km s$^{-1}$ to 50 km s$^{-1}$ is likely associated with the SNR. The ISM is dominated by the H{\sc i} gas and the mass of the molecular gas probed by CO corresponds to about 4 \% of the H{\sc i} gas. This association is supported by the morphological signs of the interaction between CO/H{\sc i} and the SNR, such as cometary-tailed and shell-like shapes, in 20 km s$^{-1}$ to 30 km s$^{-1}$. The visual extinction in the south of the SNR corresponding the ISM protons lends another support for the association.

\item The total ISM protons show a good spatial correspondence with the TeV $\gamma$-rays in azimuthal distribution. This provides a third case of such correspondence next to the SNR RX~J1713.7$-$3946 and HESS~J1713$-$347, a necessary condition for a hadronic component of the $\gamma$-ray emission. The total CR proton energy is estimated to be 10$^{48}$ erg from the $\gamma$-rays measured by H.E.S.S. and the average ISM proton density 100 cm$^{-3}$ derived from the present study. The CR energy is a similar value obtained for RX~J1713.7$-$3946 and is 0.1 \% of the total kinetic energy of a SNe.

\item The large velocity splitting of the H{\sc i} gas is likely due to the expansion of a few supershells driven by star clusters but is not due to the Galactic rotation. The velocity range from 0 km s$^{-1}$ to 60 km s$^{-1}$ corresponds to that of the inter-arm features between the local arm and the Perseus arm. We find a possible parent H{\sc i}/CO cloud where the high mass stellar projenitor of the SNR was formed. This cloud located toward the interface of a few supershells and has a mass of $10^5$ $M_\odot$ in H{\sc i}.
\end{enumerate}

\section*{APPENDIX: IDENTIFICATION OF THE SUPERSHELL GS~265$-$04$+$35}\label{sec:app}
In order to identify the supershell GS~265$-$04$+$35, we used following three steps;

\begin{enumerate}
\item {\it{Searching for a central velocity of H{\sc i} supershell}}\\
Investigating the H{\sc i} profile toward the supershell and searching for the velocity with minimum H{\sc i} intensity. We defined that the velocity is the central velocity of an expanding motion \citep[c.f.,][]{2002ApJ...578..176M}. The front and rear walls were also identified (Figure \ref{figa1}). The expanding velocity was estimated to be $\sim9$ km s$^{-1}$ as a difference between the central and front wall velocities (see also Table \ref{tab2}).

\item {\it{Estimating a geometric center of the H{\sc i} supershell}}\\
In the H{\sc i} map of the central velocity, we estimated a radial distance of the supershell as defined at peak H{\sc i} intensity every 15 degree for an assumed geometric center. We circulated only the radial distance with the peak H{\sc i} intensity of 30 K or higher. We minimized the dispersion of radial distances in azimuth angles, which gives the geometric center to be ($l$, $b$) = (264\fdg9, $-3\fdg6$) (see also Table \ref{tab2}).

\item {\it{Estimating the averaged radii and compared these with the model}}\\
We calculated the averaged radii from the geometric center for each velocity as shown in Figures \ref{figa2}a--\ref{figa2}c, and compared these with a model of expanding shell (Figures \ref{figa2}d--\ref{figa2}f). 
\end{enumerate}

Finally, we confirmed that GS~265$-$04$+$35 is the expanding H{\sc i} supershell.

\acknowledgments
{\footnotesize{The NANTEN project is based on a mutual agreement between Nagoya University and the Carnegie Institution of Washington (CIW). We greatly appreciate the hospitality of all the staff members of the Las Campanas Observatory of CIW. We are thankful to many Japanese public donors and companies who contributed to the realization of the project. This study was financially supported by Grants-in-Aid for Scientific Research (KAKENHI) of the Japanese society for the Promotion of Science (JSPS, grant Nos. 12J10082, 24224005, 25287035, and 15H05694). This work also was supported by ``Building of Consortia for the Development of Human Resources in Science and Technology'' of Ministry of Education, Culture, Sports, Science and Technology (MEXT, grant No. 01-M1-0305).}}

\software{MIRIAD \citep{1995ASPC...77..433S}}


\begin{thebibliography}{99}
\bibitem[Acero et al.(2013)]{2013A&A...551A...7A} Acero, F., Gallant, Y., Ballet, J., Renaud, M., \& Terrier, R.\ 2013, \aap, 551, A7 
\bibitem[Aharonian et al.(2004)]{2004Natur.432...75A} Aharonian, F.~A., Akhperjanian, A.~G., Aye, K.-M., et al.\ 2004, \nat, 432, 75 
\bibitem[Aharonian et al.(2005)]{2005A&A...437L...7A} Aharonian, F., Akhperjanian, A.~G., Bazer-Bachi, A.~R., et al.\ 2005, \aap, 437, L7 
\bibitem[Aharonian et al.(2006)]{2006A&A...449..223A} Aharonian, F., Akhperjanian, A.~G., Bazer-Bachi, A.~R., et al.\ 2006, \aap, 449, 223 
\bibitem[Aharonian et al.(2007a)]{2007A&A...464..235A} Aharonian, F., Akhperjanian, A.~G., Bazer-Bachi, A.~R., et al.\ 2007, \aap, 464, 235 
\bibitem[Aharonian et al.(2007b)]{2007ApJ...661..236A} Aharonian, F., Akhperjanian, A.~G., Bazer-Bachi, A.~R., et al.\ 2007, \apj, 661, 236 
\bibitem[Aharonian et al.(2009)]{2009ApJ...692.1500A} Aharonian, F., Akhperjanian, A.~G., de Almeida, U.~B., et al.\ 2009, \apj, 692, 1500 
\bibitem[Allen et al.(2015)]{2015ApJ...798...82A} Allen, G.~E., Chow, K., DeLaney, T., et al.\ 2015, \apj, 798, 82
\bibitem[Arnal \& Corti(2007)]{2007AA...476..255A} Arnal, E.~M., \& Corti, M.\ 2007, \aap, 476, 255 
\bibitem[Aschenbach(1998)]{1998Natur.396..141A} Aschenbach, B.\ 1998, \nat, 396, 141 
\bibitem[Aschenbach et al.(1999)]{1999A&A...350..997A} Aschenbach, B., Iyudin, A.~F., \& Sch{\"o}nfelder, V.\ 1999, \aap, 350, 997 
\bibitem[Bamba et al.(2005)]{2005ApJ...632..294B} Bamba, A., Yamazaki, R., \& Hiraga, J.~S.\ 2005, \apj, 632, 294 
\bibitem[Bell(1978)]{1978MNRAS.182..147B} Bell, A.~R.\ 1978, \mnras, 182, 147 
\bibitem[Bertsch et al.(1993)]{1993ApJ...416..587B} Bertsch, D.~L., Dame, T.~M., Fichtel, C.~E., et al.\ 1993, \apj, 416, 587 
\bibitem[Brand \& Blitz(1993)]{1993A&A...275...67B} Brand, J., \& Blitz, L.\ 1993, \aap, 275, 67 
\bibitem[Blandford \& Ostriker(1978)]{1978ApJ...221L..29B} Blandford, R.~D., \& Ostriker, J.~P.\ 1978, \apjl, 221, L29 
\bibitem[Cha et al.(1999)]{1999ApJ...515L..25C} Cha, A.~N., Sembach, K.~R., \& Danks, A.~C.\ 1999, \apjl, 515, L25 
\bibitem[Chen \& Gehrels(1999)]{1999ApJ...514L.103C} Chen, W., \& Gehrels, N.\ 1999, \apjl, 514, L103 
\bibitem[Dawson et al.(2008a)]{2008MNRAS.387...31D} Dawson, J.~R., Mizuno, N., Onishi, T., McClure-Griffiths, N.~M., \& Fukui, Y.\ 2008, \mnras, 387, 31 
\bibitem[Dawson et al.(2008b)]{2008PASJ...60.1297D} Dawson, J.~R., Kawamura, A., Mizuno, N., Onishi, T., \& Fukui, Y.\ 2008, \pasj, 60, 1297 
\bibitem[Dawson et al.(2011a)]{2011ApJ...728..127D} Dawson, J.~R., McClure-Griffiths, N.~M., Kawamura, A., et al.\ 2011, \apj, 728, 127 
\bibitem[Dawson et al.(2011b)]{2011ApJ...741...85D} Dawson, J.~R., McClure-Griffiths, N.~M., Dickey, J.~M., \& Fukui, Y.\ 2011, \apj, 741, 85 
\bibitem[Dawson et al.(2015)]{2015ApJ...799...64D} Dawson, J.~R., Ntormousi, E., Fukui, Y., Hayakawa, T., \& Fierlinger, K.\ 2015, \apj, 799, 64 
\bibitem[Dickey \& Lockman(1990)]{1990ARA&A..28..215D} Dickey, J.~M., \& Lockman, F.~J.\ 1990, \araa, 28, 215 
\bibitem[Dobashi et al.(2005)]{2005PASJ...57S...1D} Dobashi, K., Uehara, H., Kandori, R., et al.\ 2005, \pasj, 57, S1 
\bibitem[Ellison et al.(2010)]{2010ApJ...712..287E} Ellison, D.~C., Patnaude, D.~J., Slane, P., \& Raymond, J.\ 2010, \apj, 712, 287 
\bibitem[Fukuda et al.(2014)]{2014ApJ...788...94F} Fukuda, T., Yoshiike, S., Sano, H., et al.\ 2014, \apj, 788, 94 
\bibitem[Fukui et al.(1999)]{1999PASJ...51..751F} Fukui, Y., Onishi, T., Abe, R., et al.\ 1999, \pasj, 51, 751 
\bibitem[Fukui et al.(2003)]{2003PASJ...55L..61F} {Fukui, Y., Moriguchi, Y., Tamura, K., et al.\ 2003, \pasj, 55, L61}
\bibitem[Fukui et al.(2012)]{2012ApJ...746...82F} Fukui, Y., Sano, H., Sato, J., et al.\ 2012, \apj, 746, 82 
\bibitem[Fukui et al.(2014)]{2014ApJ...796...59F} Fukui, Y., Okamoto, R., Kaji, R., et al.\ 2014, \apj, 796, 59 
\bibitem[Fukui et al.(2015)]{2015ApJ...798....6F} Fukui, Y., Torii, K., Onishi, T., et al.\ 2015, \apj, 798, 6 
\bibitem[Gabici et al.(2007)]{2007Ap&SS.309..365G} Gabici, S., Aharonian, F.~A., \& Blasi, P.\ 2007, \apss, 309, 365 
\bibitem[Gabici \& Aharonian(2014)]{2014MNRAS.445L..70G} {Gabici, S., \& Aharonian, F.~A.\ 2014, \mnras, 445, L70}
\bibitem[Giuliani et al.(2010)]{2010A&A...516L..11G} Giuliani, A., Tavani, M., Bulgarelli, A., et al.\ 2010, \aap, 516, L11
\bibitem[Giuliani et al.(2011)]{2011ApJ...742L..30G} Giuliani, A., Cardillo, M., Tavani, M., et al.\ 2011, \apjl, 742, L30 
\bibitem[Hess \& Steinmaurer(1935)]{1935Natur.135..617H} Hess, V.~F., \& Steinmaurer, R.\ 1935, \nat, 135, 617 
\bibitem[H.E.S.S.~Collaboration et al.(2011)]{2011A&A...531A..81H} H.E.S.S.~Collaboration, Abramowski, A., Acero, F., et al.\ 2011, \aap, 531, A81 
\bibitem[H.E.S.S.~Collaboration et al.(2016a)]{2016arXiv160104461H} H.E.S.S.~Collaboration, Abramowski, A., Aharonian, F., et al.\ 2016a, arXiv:1601.04461 
\bibitem[H.E.S.S.~Collaboration et al.(2016b)]{2016arXiv160908671H} H.E.S.S.~Collaboration, Abdalla, H., Abdalla, H., et al.\ 2016a, arXiv:1609.08671 
\bibitem[H.E.S.S.~Collaboration et al.(2016c)]{2016arXiv161101863H} H.E.S.S.~Collaboration, Abdalla, H., Abramowski, A., et al.\ 2016b, arXiv:1611.01863
\bibitem[Inoue et al.(2012)]{2012ApJ...744...71I} Inoue, T., Yamazaki, R., Inutsuka, S.-i., \& Fukui, Y.\ 2012, \apj, 744, 71 
\bibitem[Iyudin et al.(2005)]{2005A&A...429..225I} Iyudin, A.~F., Aschenbach, B., Becker, W., Dennerl, K., \& Haberl, F.\ 2005, \aap, 429, 225 
\bibitem[Iyudin et al.(2007)]{2007ESASP.622...91I} Iyudin, A.~F., Aschenbach, V., Burwitz, V., et al.\ 2007, The Obscured Universe.~Proceedings of the VI INTEGRAL Workshop, 622, 91
\bibitem[Katsuda et al.(2008)]{2008ApJ...678L..35K} Katsuda, S., Tsunemi, H., \& Mori, K.\ 2008, \apjl, 678, L35 
\bibitem[Kalberla et al.(2005)]{2005A&A...440..775K} Kalberla, P.~M.~W., Burton, W.~B., Hartmann, D., et al.\ 2005, \aap, 440, 775 
\bibitem[Liseau et al.(1992)]{1992A&A...265..577L} Liseau, R., Lorenzetti, D., Nisini, B., Spinoglio, L., \& Moneti, A.\ 1992, \aap, 265, 577 
\bibitem[Matsunaga et al.(2001)]{2001PASJ...53.1003M} Matsunaga, K., Mizuno, N., Moriguchi, Y., et al.\ 2001, \pasj, 53, 1003 
\bibitem[Maxted et al.(2017)]{2017MNRAS...submitted} Maxted, N., Burton, M., \& Braiding, C., et al.\ 2017, submitted to \mnras
\bibitem[May et al.(1988)]{1988A&AS...73...51M} May, J., Murphy, D.~C., \& Thaddeus, P.\ 1988, \aaps, 73, 51 
\bibitem[McClure-Griffiths et al.(2002)]{2002ApJ...578..176M} McClure-Griffiths, N.~M., Dickey, J.~M., Gaensler, B.~M., \& Green, A.~J.\ 2002, \apj, 578, 176 
\bibitem[McClure-Griffiths et al.(2005)]{2005ApJS..158..178M} McClure-Griffiths, N.~M., Dickey, J.~M., Gaensler, B.~M., et al.\ 2005, \apjs, 158, 178 
\bibitem[Mereghetti(2001)]{2001ApJ...548L.213M} Mereghetti, S.\ 2001, \apjl, 548, L213 
\bibitem[Moriguchi et al.(2001)]{2001PASJ...53.1025M} Moriguchi, Y., Yamaguchi, N., Onishi, T., Mizuno, A., \& Fukui, Y.\ 2001, \pasj, 53, 1025
\bibitem[Moriguchi et al.(2005)]{2005ApJ...631..947M} {Moriguchi, Y., Tamura, K., Tawara, Y., et al.\ 2005, \apj, 631, 947}
\bibitem[Murphy \& May(1991)]{1991A&A...247..202M} Murphy, D.~C., \& May, J.\ 1991, \aap, 247, 202 
\bibitem[Okamoto et al.(2017)]{2017ApJ...838..132O} Okamoto, R., Yamamoto, H., Tachihara, K., et al.\ 2017, \apj, 838, 132
\bibitem[Pannuti et al.(2010)]{2010ApJ...721.1492P} Pannuti, T.~G., Allen, G.~E., Filipovi{\'c}, M.~D., et al.\ 2010, \apj, 721, 1492 
\bibitem[Pavlov et al.(2001)]{2001ApJ...559L.131P} Pavlov, G.~G., Sanwal, D., K{\i}z{\i}ltan, B., \& Garmire, G.~P.\ 2001, \apjl, 559, L131 
\bibitem[Planck Collaboration et al.(2014)]{2014A&A...571A..11P} Planck Collaboration, Abergel, A., Ade, P.~A.~R., et al.\ 2014, \aap, 571, A11 
\bibitem[Roy et al.(2013)]{2013ApJ...763...55R} Roy, A., Martin, P.~G., Polychroni, D., et al.\ 2013, \apj, 763, 55 
\bibitem[Sano et al.(2010)]{2010ApJ...724...59S} Sano, H., Sato, J., Horachi, H., et al.\ 2010, \apj, 724, 59 
\bibitem[Sano et al.(2013)]{2013ApJ...778...59S} Sano, H., Tanaka, T., Torii, K., et al.\ 2013, \apj, 778, 59 
\bibitem[Sano et al.(2015)]{2015ApJ...799..175S} Sano, H., Fukuda, T., Yoshiike, S., et al.\ 2015, \apj, 799, 175 
\bibitem[Sano et al.(2017)]{sano2017inprep} Sano, H., et al.\ 2017, in preparation
\bibitem[Sault et al.(1995)]{1995ASPC...77..433S} Sault, R.~J., Teuben, P.~J., \& Wright, M.~C.~H.\ 1995, Astronomical Data Analysis Software and Systems IV, 77, 433
\bibitem[Slane et al.(2001a)]{2001ApJ...548..814S} Slane, P., Hughes, J.~P., Edgar, R.~J., et al.\ 2001, \apj, 548, 814 
\bibitem[Slane et al.(2001b)]{2001AIPC..565..403S} Slane, P., Hughes, J.~P., Edgar, R.~J., et al.\ 2001, Young Supernova Remnants, 565, 403 
\bibitem[Suad et al.(2014)]{2014AA...564A.116S} Suad, L.~A., Caiafa, C.~F., Arnal, E.~M., \& Cichowolski, S.\ 2014, \aap, 564, A116 
\bibitem[Takeda et al.(2016)]{2016PASJ...68S..10T} Takeda, S., Bamba, A., Terada, Y., et al.\ 2016, \pasj, 68, S10
\bibitem[Tsunemi et al.(2000)]{2000PASJ...52..887T} Tsunemi, H., Miyata, E., Aschenbach, B., Hiraga, J., \& Akutsu, D.\ 2000, \pasj, 52, 887 
\bibitem[Vall{\'e}e(2008)]{2008AJ....135.1301V} Vall{\'e}e, J.~P.\ 2008, \aj, 135, 1301-1310 
\bibitem[Yamaguchi et al.(1999a)]{1999PASJ...51..765Y} Yamaguchi, N., Mizuno, N., Moriguchi, Y., et al.\ 1999a, \pasj, 51, 765
\bibitem[Yamaguchi et al.(1999b)]{1999PASJ...51..775Y} Yamaguchi, N., Mizuno, N., Saito, H., et al.\ 1999b, \pasj, 51, 775 
\bibitem[Yoshiike et al.(2013)]{2013ApJ...768..179Y} Yoshiike, S., Fukuda, T., Sano, H., et al.\ 2013, \apj, 768, 179
\bibitem[Yoshiike et al.(2017)]{2017ApJ...submitted} Yoshiike, S., Fukuda, T., Sano, H., et al.\ 2017, \apj submitted
\bibitem[Zirakashvili \& Aharonian(2010)]{2010ApJ...708..965Z} {Zirakashvili, V.~N., \& Aharonian, F.~A.\ 2010, \apj, 708, 965}
\end{thebibliography}
\end{document}